\documentclass[journal]{IEEEtran}
\usepackage[utf8]{inputenc}
\usepackage[T1]{fontenc}
\usepackage[brazil,english]{babel}
\usepackage{graphicx}
\usepackage{amssymb}
\usepackage{graphics}
\usepackage{bibentry}
\usepackage{epsfig}
\usepackage{booktabs}
\usepackage{lscape}
\usepackage{enumitem}
\usepackage{float}
\usepackage{multirow}
\usepackage{amsmath}
\usepackage{textcomp}
\usepackage{adjustbox}
\usepackage{breqn}
\usepackage{cite}
\usepackage{multirow}
\usepackage[usenames, dvipsnames]{color}
\usepackage{makecell}
\usepackage[ruled,vlined]{algorithm2e}
\usepackage{algpseudocode}
\usepackage{tabu}
\usepackage{xcolor}
\usepackage{colortbl}
\usepackage[normalem]{ulem}
\useunder{\uline}{\ul}{}

\ifCLASSINFOpdf
\else
\fi
\begin{document}
\title{Hybrid deep learning-based strategy for hepatocellular carcinoma cancer grade classification 
}



\makeatletter
\newcommand{\linebreakand}{%
  \end{@IEEEauthorhalign}
  \hfill\mbox{}\par
 \mbox{}\hfill\begin{@IEEEauthorhalign}mpq
}
\makeatother

\author{Ajinkya~Deshpande,~Deep~Gupta~\IEEEmembership{Senior~Member,~IEEE}, 
~Ankit Bhurane~\IEEEmembership{Member,~IEEE}, ~Nisha Meshram, ~Sneha Singh~\IEEEmembership{Member,~IEEE}, Petia  Radeva  ~\IEEEmembership{Fellow,~IAPR} 
\thanks{A. Deshpande, D. Gupta and A Bhurane are with the Department
of Electronics and Communication Engineering, Visvesvaraya National Institute of Technology, Nagpur, MH, India (e-mail: 39ajinkya@gmail.com, deepgupta@ece.vnit.ac.in, ankitbhurane@ece.vnit.ac.in)}
\thanks{N. Meshram is with the Department
of Pathology, AIIMS Nagpur, MH, India (e-mail: drnisha@aiimsnagpur.edu.in.)}
\thanks{Sneha Singh is with the School of Computing \& Electrical Engineering, the Indian Institute of Technology Mandi, India (e-mail:sneha@iitmandi.ac.in)}
\thanks{P. Radeva is with the Department of Mathematics and Informatics, Universitat de Barcelona, 08007,  (Barcelona), Spain (e-mail:petia.ivanova@ub.edu).}}

\maketitle

\begin{abstract}

Hepatocellular carcinoma (HCC) is a common type of liver cancer whose early-stage diagnosis is a challenge. This is mainly done by a manual assessment of hematoxylin and eosin-stained whole slide images, which is a time-consuming process with high variability in the decision-making. In order to achieve accurate detection of HCC, we propose a hybrid deep learning architecture that leverages the advantages of fine-tuning selective layers of convolutional neural network (CNN) alongside a deep classifier comprising a sequence of fully connected layers, i.e., Artificial Neural Network (ANN).
In order to validate the method, this study uses a publicly available The Cancer Genome Atlas Hepatocellular Carcinoma (TCGA-LIHC)\cite{tcga} database (n=491) for model development and proprietary database of Kasturba Gandhi Medical College (KMC), India \cite{LiverNet} for validation.
The proposed hybrid model with ResNet50-based feature extractor provided the sensitivity, specificity, F1-score, accuracy, and AUC of 100.00\%, 100.00\%, 100.00\%, 100.00\%, and 1.00, respectively on the TCGA database. It was contrasted with eight different state-of-the-art models. On the KMC database, EfficientNetb3 resulted in the optimal choice of the feature extractor giving sensitivity, specificity, F1-score, accuracy, and AUC of 96.97$\pm$0.83, 98.85$\pm$0.24, 96.71$\pm$0.66, 96.71$\pm$0.66, and 0.99$\pm$0.01, respectively. The proposed hybrid models showed improvement in accuracy of 2\% and 4\% over the pre-trained models in TCGA-LIHC and KMC databases.

\end{abstract}


\section{Introduction}
\IEEEPARstart{G}{lobally}, over 700,000 people die every year from liver cancer, making it the third most common cause of cancer-related death \cite{El-Serag2007-kz}\cite{american}. The hepatocellular carcinoma (HCC) is the most common type of liver cancer accounting for almost 80\% of all the primary liver cancer cases\cite{burden_cancer, li2019staged}. A major spread of the HCC has been observed in Asian and African countries. Pathologists consider histopathology imaging as a gold standard to identify HCC \cite{Chen2022-vy}. However, due to the diversity in tumor shape, tissue sizes and staining procedure, manual assessment of such histopathology images is often a challenging and error-prone task \cite{Raab2005-im}\cite{Elmore2015-fd}\cite{Nakhleh2006-fb}.

With advances in artificial intelligence (AI), the landscape of medical diagnosis has completely changed, and it has helped over the years the medical community in providing patients with accurate and inexpensive computer-aided diagnostic solutions for the betterment of their health \cite{greenspan2016guest}. In the domain of liver cancer detection, the AI-based convolutional neural network (CNN) algorithms have contributed to a great extent to the prediction of cancer types \cite{li2019staged} \cite{LiverNet}\cite{global}\cite{Chen2020}\cite{Chen2022-vy} replacing traditional techniques such as SVM\cite{cortes1995support}. In the past decade, several CNN-based architectures have been proposed that either focus on deepening the network such as AlexNet\cite{NIPS2012_c399862d} and VGG\cite{VGG16}, overcoming the vanishing gradient issue by adding the residual connections, ResNet \cite{ResNet50} architectures and DenseNet \cite{DenseNet121} or applying compound scaling methodology that demonstrates dependency on width, depth and resolution of CNN through EfficientNet architectures \cite{EfficientNet}. These models have been used as state-of-the-art for a long time and possess special features. Most of the recent studies used this CNN architecture as their backbone models. Unlike the conventional machine learning (ML)--based algorithms, the CNNs do not require handcrafted features for training. The architecture of CNN is responsible for both feature extraction and classification, eventually achieving better performance than traditional methods\cite{BreastNet}\cite{9813383} and providing an end-to-end solution. This is main reason for their popularity in performing medical image classification \cite{DBLP:ml_cv}.

\begin{figure}[!t]
\centering
\scalebox{0.45}{\includegraphics[width=7.4in, 
		]{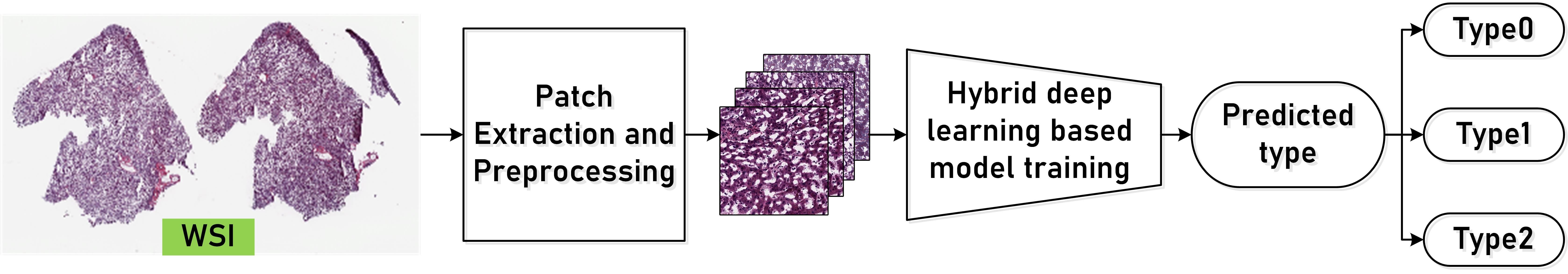}}
\caption{Block diagram showing overall workflow of the experimentation} \label{Fig: overall_workflow}
\end{figure}

Recently, Sun et al.\cite{global} used a ResNet50 
 to perform the liver cancer classification using global image-based labels. The authors proposed patch extraction and feature selection mechanisms. Chen et al.\cite{Chen2020} used the InceptionV3  to perform the patch-level liver cancer classification on the TCGA dataset. Atresh et al.\cite{LiverNet} proposed a new LiverNet architecture, which consisted of a convolutional block attention module block (CBAM), atrous convolution spatial pyramid pooling blocks, and hypercolumn technique, which reduces model parameters as well as makes the model more robust and precise. The authors claimed an overall liver cancer detection accuracy of 97.72\%. In another study, Chen et al.\cite{Chen2022-vy} studied the performance of ResNet50, ResNet50\_CBAM, SENet, and SKNet on purely, moderately and well-differentiated histopathology images. SENet and ResNet50 were found to be the best-performing models on poorly and well-differentiated types, respectively. Another study \cite{9921287} proposes prior segmentation for better classification accuracy. Luo et al. \cite{10298268} employed a deformable convolution-guided attention block and deep adaptive feature fusion, and proposed DCA-DAFFNet for laryngeal tumor grading achieving an accuracy of 90.78\%. Similarly, few more recent studies used the CNN-based algorithms for histopathology image classification \cite{app10103359}\cite{SHALLU2018247} \cite{BreastNet}\cite{sari2018unsupervised}. 

All these studies used diverse CNN-based architecture for liver cancer classification. Most of these studies relied on transfer learning, particularly fine-tuning, which is its most widely used form. In transfer learning, the model trained on a diversified and huge database with multiple classes is used to train on a smaller dataset with a limited number of classes\cite{hussain2018study}\cite{TaloM} \cite{TALO2019176} \cite{10491332}. Similarly, in fine-tuning,  selective top layers of feature extractor together with the classifier are made trainable\cite{app10103359}\cite{DBLP:journals/corr/YosinskiCBL14}\cite{Sarhan2022}\cite{SHALLU2018247}\cite{7426826}. A recent study by Talo et al.\cite{TALO2019176} suggested that if some of the top layers of the feature extractor are kept trainable along with the classifier, then better classification accuracy can be achieved. 

Motivated by the concept of transfer learning, and observing that keeping selective top feature extractor layers along with classifier trainable can boost overall classification performance\cite{app10103359}, we developed a hybrid CNN-based architecture for liver cancer classification. We hypothesized that the proposed hybrid deep learning-based architecture performs better than the individual pre-trained models. Hence, our contributions here are  as follows:
\begin{enumerate}
    \item We propose a deep classifier designed with gradual feature reduction to ensure proper mapping to the output label space. This also helps to develop an optimum end-to-end solution for histopathological image classification.
    \item We applied a fine-tuning-powered model to address the issues in cross-domain learning. In the proposed work, the effect of transferring the learning to the medical domain and selectively tuning the model for new characterization is studied.
    \item This study also demonstrates the importance of proper image pre-processing and training methodology for better loss convergence, achieving optimal results and developing a robust model for carcinoma detection in histopathological images.
\end{enumerate}



The rest of the paper is organized as follows. Section 2 discusses the details of the dataset, the preprocessing, the proposed model architecture, the performance metrics and the training methodology. Section 3 describes the experimental details within the validation along with a detailed comparative study. In Section 4, a detailed discussion is given of the experimental results obtained, followed by the conclusions in Section 5. The overall workflow for the experimentation is shown in Fig. \ref{Fig: overall_workflow}.

\section{Methods}

\subsection{Datasets and Pre-processing}

This study uses a publicly available cancer genome atlas program liver hepatocellular carcinoma (TCGA-LIHC) database of 491 whole slide images \cite{tcga} to train our proposed hybrid deep learning-based model. The 491 slides were divided into three types of liver HCC tumor namely solid tissue normal as type 0, primary tumor as type 1, and recurrent tumor as type 2. There were 89 solid tissue normal slides, 399 primary tumor slides, and 3 recurrent tumor slides. In a deep learning-based approach for classification, directly feeding the whole slide images of huge dimensions to CNN is nearly impractical as it increases computations and model parameters by a large amount. Downsampling or reshaping the histopathology images to lower dimensions leads to severe information loss. Therefore, we follow a patch-based approach \cite{global}, where each WSI slide is tiled into patches of size 1024$\times$1024. While performing the patch extraction, we also intended to maintain sufficient visibility of tissues in the patches. This was ensured by maintaining the mean intensity value of 200 and the standard deviation of 60. An empirical selection process is performed to maintain a clear visualization of tissues in the patches and reduce the class imbalance in the database. 
Finally, this process produced 813, 893, and 680 valid patches of type 0, type 1, and type 2, respectively, which are considered for developing the hybrid deep learning-based model. 

\begin{figure}[!t]
\centering
\scalebox{0.25}{\includegraphics[width=7.4in, 
		]{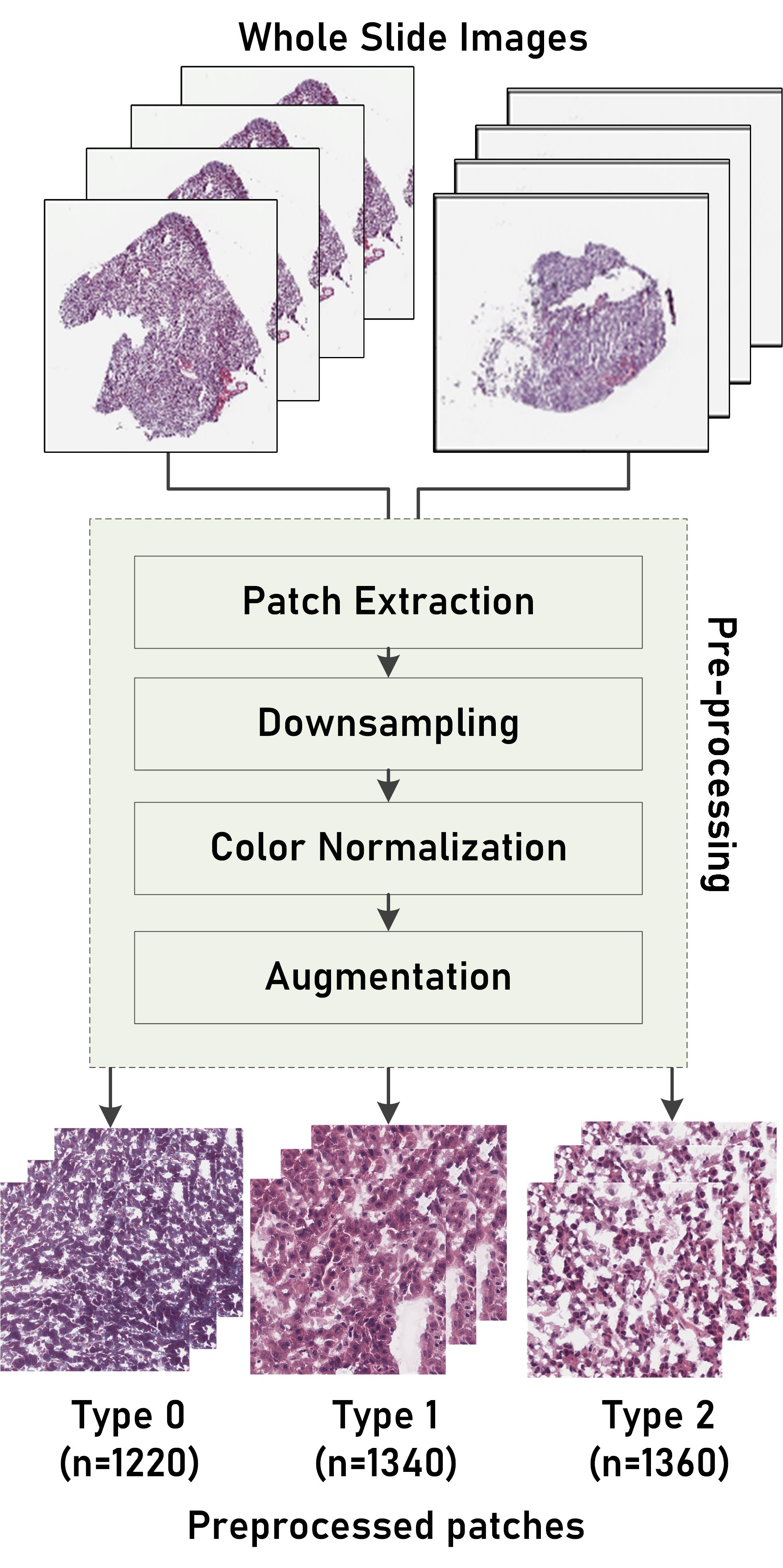}}
\caption{TCGA-LIHC liver HCC database\cite{tcga}. This diagram describes the preprocessing for TCGA dataset preparation, which involves patch extraction from the whole slide image, downsampling to balance the dataset, and data augmentation. After data preprocessing, the number of images obtained are mentioned at the bottom.} \label{Fig1}
\end{figure}

The color normalization is another important step in the pre-processing of histology images because of the variation of staining intensity of hematoxylin and eosin used for staining purposes in the histology images by the pathologist and may vary according to the practitioner and physical conditions. Thus,  color variations can significantly impact model training, because models are color-sensitive, In the proposed study, we used the color normalization technique proposed by Macenko et al.\cite{color_norm} which presents an algorithm that finds the correct stain vectors for the image after converting image pixels to optical density space automatically and then performs color deconvolution.

Deep learning models require a diverse and large dataset to build robust and flexible networks. Augmentation is a technique that generates variants of existing data, ultimately increasing the dataset. This makes data augmentation an important step in preprocessing. Over the normalized patches, we applied random vertical and horizontal flips and obtained 1220, 1340, and 1360 patches of type 0, type 1, and type 2 respectively. The complete pre-processing stage is depicted in Fig. \ref{Fig1}.

 This study also uses another cancer database of liver HCC collected from Kasturba Gandhi Medical College, India \cite{LiverNet}. The KMC database has four types of liver HCC tumors such as type 0, type 1, type 2, and type 3 having 719, 799, 776, and 711 image patches, respectively, each of size 224$\times$224$\times$3. The KMC database received\cite{LiverNet}, was originally pre-processed and, therefore, only the augmentation step is performed during the training of the proposed model. Moreover, to verify the proposed hybrid learning strategy for different types of cancer disease, another publicly available colon cancer database of histopathology images is collected from Kaggle \cite{colon2019}. The colon database consists of two classes, adenocarcinoma and normal, with a total of 5000 patches per type of size 768$\times$768$\times$3. Samples of all three datasets are shown in Fig. \ref{Fig3}

\begin{figure}[!t]
\centering
\scalebox{0.25}{\includegraphics[width=10in, height=9in
		]{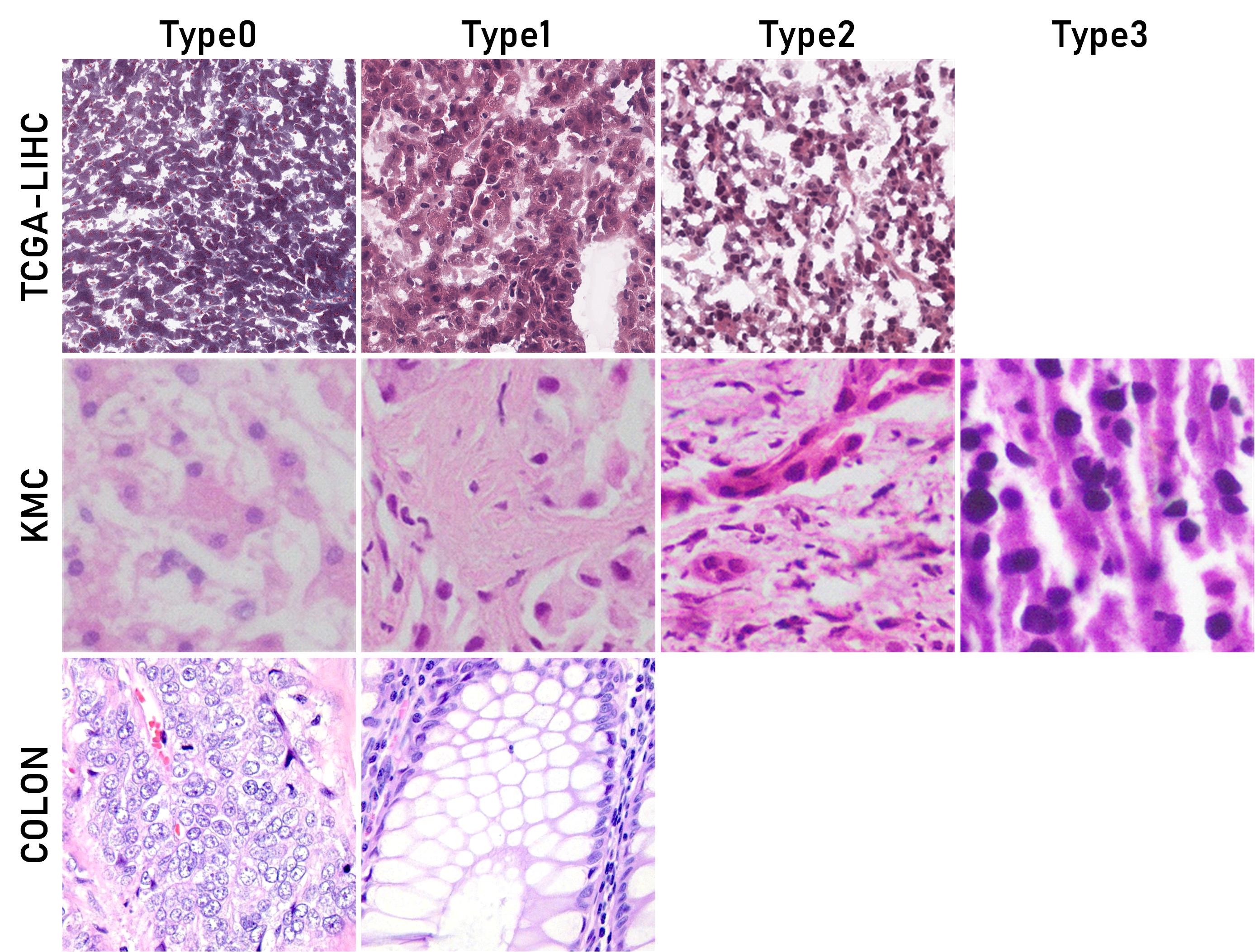}}
\caption{TCGA-LIHC, KMC and COLON datasets are comprised of 3, 4 and 2 types respectively. This figure shows demo patches available in the respective dataset.} \label{Fig3}
\end{figure}

\subsection{Proposed model architecture}
The transfer learning property of NNs allows to apply models that have been trained on large diverse datasets (i.e., “pre-trained models”) to smaller datasets from completely different domains in a more efficient manner \cite{app10103359, DBLP:journals/corr/YosinskiCBL14, Sarhan2022, TALO2019176, TaloM, global}. In general, the pre-trained models have two parts: a feature extractor and a classifier. In transfer learning, the classifier is modified while the feature extractor is left untouched. This is because the pre-trained models are usually trained on the Imagenet dataset with 1000 output classes, while in general, the number of output classes is limited. Therefore, only the last classification layer of the pre-trained model is re-trained, named as “base model". 

In general, the top layers of any pre-trained model's feature extractor capture the fine details that pertain solely to output classes, and therefore, such top layers are re-trained when fine-tuning the model on small custom datasets to optimize the performance of the model \cite{app10103359, DBLP:journals/corr/YosinskiCBL14},\cite{Sarhan2022}. In contrast to the top layers, the bottom layers capture the more generic features such as lines, edges, and shapes that contribute more to the classification task. Besides the feature extractor and the classifier, another important attribute of the pre-trained model is the depth of the model. There has been evidence that deeper models can improve classification performance, although this is not always the case \cite{ResNet50, DenseNet121, EfficientNet}. Considering all these aspects, we propose an architecture in which (i) the feature extractor remains unchanged except for a few top layers and (ii) a few more fully connected layers (ANN) are added before the final classification layer. This modified model, referred to as the "hybrid model", leverages the advantages of selective fine-tuning and deep classifiers. The number of additional fully connected layers is decided empirically. The performance of eight different types of pre-trained models such as ResNet50\cite{ResNet50}, VGG16\cite{VGG16}, EfficientNetb0\cite{EfficientNet}, EfficientNetb1\cite{EfficientNet}, EfficientNetb2\cite{EfficientNet}, EfficientNetb3\cite{EfficientNet}, EfficientNetb4\cite{EfficientNet}, DenseNet121\cite{DenseNet121} is compared while designing the hybrid model. Fig. \ref{fig:4} shows the architecture of the base and hybrid models.

\begin{figure}[t!]
    \centering
    \scriptsize
    \scalebox{0.45}{\includegraphics[width=7.4in, 
		]{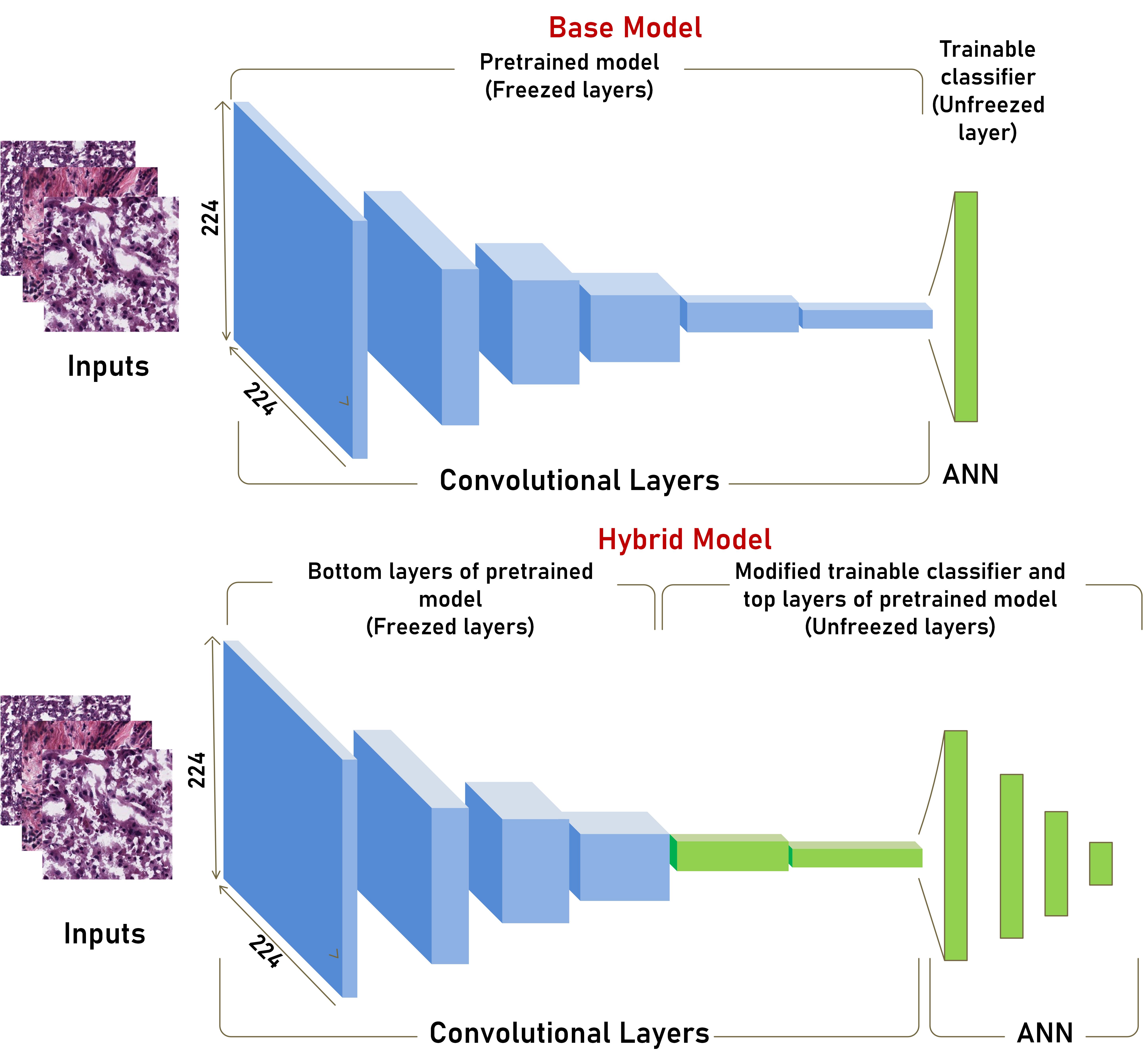}}
    \caption{i) Base model: All of the convolutional layers in pre-trained model are frozen and the last fully connected layer is replaced and kept trainable. ii) Hybrid model: Bottom layers of the convolutional layers of the pre-trained model are frozen. A shallow classifier is replaced with a deep classifier. The classifier and selective top layers of the pre-trained model are kept trainable.}
    \label{fig:4}
\end{figure}

The overall architecture of the proposed model is comprised of two stages: (i) offline system and (ii) online system as shown in Fig. \ref{Fig5}.

\subsection{Training}
  The image patches are resized to 224$\times$224$\times$3 pixel dimensions to maintain uniformity across datasets and allow faster and more precise training. The offline system uses the training dataset and performs the 5-fold stratified cross-validation to train the proposed model. In 5-fold cross-validation, the training data is divided into five equal non-overlapping parts. The model is trained on the four parts and validated on the remaining part of the training dataset. Dynamic augmentation is performed on these four training parts of the data with random rotation on the TCGA-LIHC dataset, random rotation with random horizontal flip on the COLON dataset, and random rotation with random horizontal and vertical flip on the KMC dataset. Furthermore, to minimize the class imbalance, a weighted random sampler approach \cite{NEURIPS2019_9015} is followed that samples images of different classes with pre-defined weights. The weights are inversely utilized based on the number of classes in the training dataset. Both the dynamic augmentation and weighted random sampler methods aid in reducing the class-imbalance effect in the dataset. This complete process is repeated five times in 5-fold cross-validation and an averaged validation performance is recorded. Table \ref{tab:Table2} shows the hyperparameters used while training the model. 

  Furthermore, cross-entropy loss $(L_{CE})$ between labels and ground truth values is used, which is defined in Eq.[\ref{eqn: 1}]
\begin{equation*}
L_{CE}= -\sum_{i=0}^{N-1}y_{i}\log(p_{i})
\tag{1}
\label{eqn: 1}
\end{equation*}
where $N$ is the number of classes, and $y_i$ and $p_i$ are the ground truth values and the predicted probabilistic values, respectively. This loss function is widely preferred over the others. The reason for this is that, if we see the curve of loss vs. probability, loss is very high when prediction is inaccurate. As a result, the gradient of loss is high, which results in faster convergence. Once the model is trained, the training weights are used to transform the test dataset into the output cancer types using the test model. The performance of the model is evaluated on the test data using five performance measures such as accuracy, sensitivity, specificity, F1-score, and area under the curve (AUC). The pre-trained models were imported from the Pytorch library and trained on Google Colab.

\section {Validation}
 In this section, we present the model training settings, performance metrics, and results obtained on the three datasets, followed by detailed discussions.
\subsection{Settings}
PyTorch framework was used for designing the model. Initially, the pre-processed image patches are divided into training (90\%) and testing datasets (10\%) as shown in Table \ref{tab:Table1}.
Cosine annealing warm restart \cite{loshchilov2017sgdr} learning rate scheduler is used to select the learning rate over all epochs. This is mathematically represented in Eq.[\ref{eqn: 2}], where, $\eta^{i}_{min}$ and $\eta^{i}_{max}$ are learning rate ranges. $T_{cur}$ indicates the number of epochs from the last restart, and $T_{i}$ indicates the number of epochs after which restart is scheduled. The model is trained for 47 epochs with early stopping criteria and a warm restart of the learning rate scheduler at every $12^{th}$ epoch($T_{i}$). In this study, an initial learning rate of 0.001($\eta^{i}_{max}$) is used:
\begin{equation*}
    \eta_{t} = \eta_{min}^{i} + \frac{1}{2}\left(\eta_{max}^{i}-\eta_{min}^{i}\right)\left(1+\cos\left(\frac{T_{cur}}{T_{i}}\pi\right)\right)
\tag{2}
\label{eqn: 2}
\end{equation*}

\begin{figure}[!t]
\centering
\scalebox{0.45}{\includegraphics[width=7.4in, 
		]{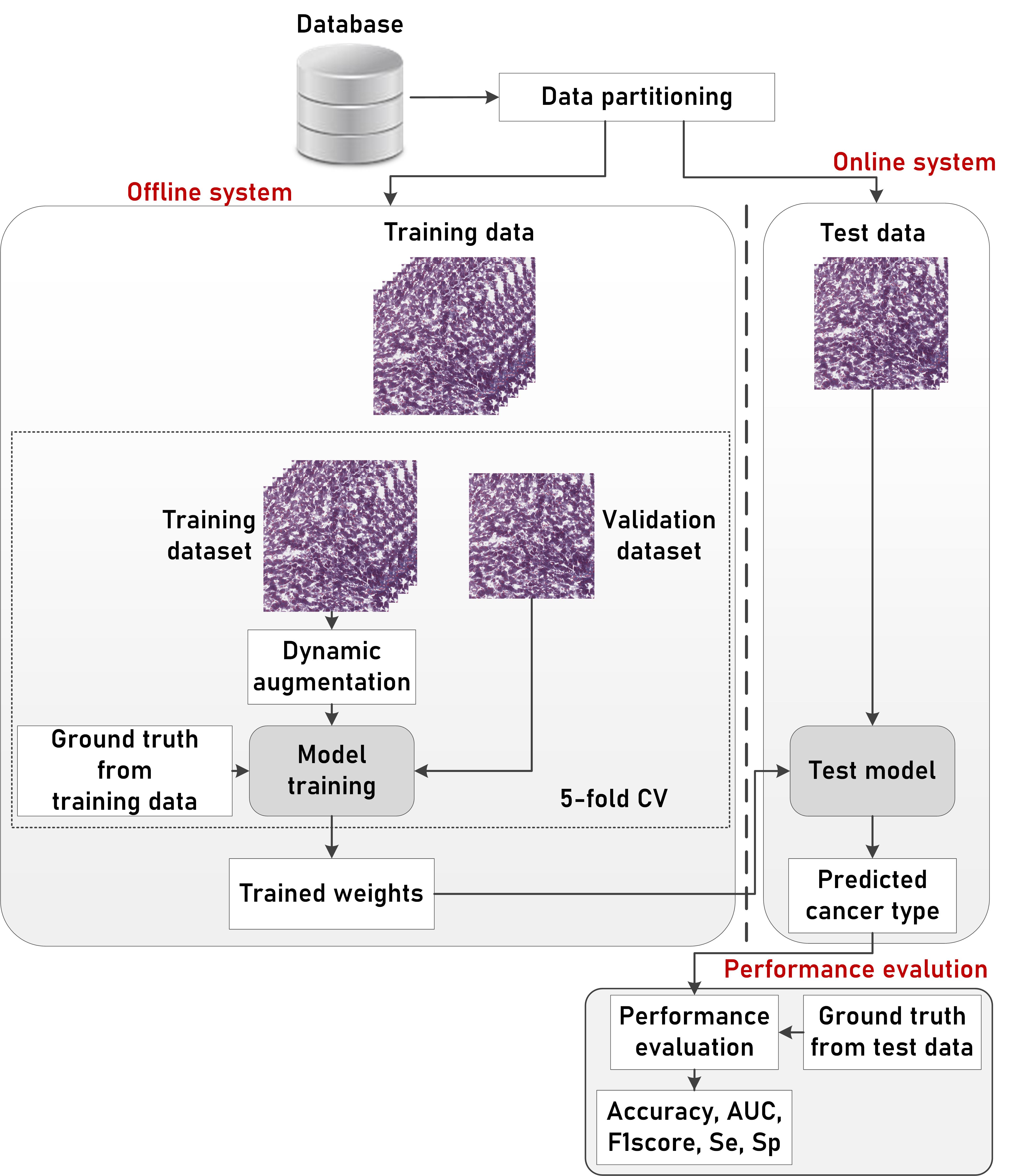}}
\caption{The workflow of the training and validation of the proposed method } \label{Fig5}
\end{figure}

\begin{table}[!ht]
    \centering
    \caption{Number of patches and their distribution}
    \label{tab:Table1}
    \begin{adjustbox}{width=0.46\textwidth}
    \begin{tabular}{lllllll} \hline
        \textbf{Dataset} & \textbf{} & \textbf{Type0} & \textbf{Type1} & \textbf{Type2} & \textbf{Type3} & \textbf{Total } \\ \hline 
        \multirow{3}{*}{TCGA-LIHC} & Train & 1098 & 1206 & 1224 & NA & 3528 \\ 
        ~ & Test & 122 & 134 & 136 & NA & 392 \\ 
        ~ &Total & 1220 & 1340 & 1360 & NA & 3920 \\ \hline
        \multirow{3}{*}{KMC} & Train & 649 & 719 & 696 & 661 & 2725 \\ 
        ~ & Test & 70 & 80 & 80 & 50 & 280 \\ 
        ~ & Total & 719 & 799 & 776 & 711 & 3005 \\ \hline
        \multirow{3}{*}{COLON} & Train  & 4500 & 4500 & NA & NA & 9000 \\ 
        ~ & Test & 500 & 500 & NA & NA & 1000 \\ 
        ~ & Total & 5000 & 5000 & NA & NA & 10000 \\ \hline
    \end{tabular}
    \end{adjustbox}
\end{table}

\begin{table}[!ht]
    \centering
    \caption{Hyperparameters used during training of models}
    \label{tab:Table2}
    \begin{tabular}{ll}
    \hline
        \textbf{Parameter} & \textbf{Value} \\ \hline
        Batch size & 64 or 32 \\ 
        Learning rate & 0.001 \\ 
        Optimizer & Adam \\ 
        Beta1 & 0.9 \\ 
        Beta2 & 0.999 \\ 
        Epochs & 47 \\ 
        Learning rate scheduler & Cosine annealing warm restart \\
        Warm Restart & 12 \\ 
        Loss & Cross entropy loss \\ 
         \hline
    \end{tabular}
\end{table}

\subsection{Performance Metrics}
The quantitative performance of the proposed hybrid model was evaluated using a standard protocol such as accuracy, sensitivity, specificity, and F1-score \cite{xia2023weakly, li2019staged}. These performance evaluation metrics were computed using a contingency table that holds true positive (TP), false positive (FP), true negative (TN), and false negative (FN) rates. Finally, receiver operating characteristics (ROC) and area under the curve (AUC) \cite{5599384} were determined.

\subsection{Results}
In this study, the proposed hybrid model is independently trained and evaluated on the TCGA-LIHC, KMC and COLON datasets as per the strategy discussed in the “Training” section. The training and validation accuracy along with the training and validation loss are plotted for all the datasets. The five performance metrics namely accuracy, sensitivity, specificity, F1-score and AUC for each model were averaged over five-folds and presented in this section along with the ROC curves and confusion matrix attributed to respective folds. 

\subsubsection{Results on the TCGA-LIHC dataset}

To investigate the training and validation performance of the proposed hybrid model, the training, validation accuracy and loss are shown in Fig. \ref{Fig: resnet50 loss} and Fig. \ref{Fig: resnet50 accuracy}, respectively. It has been observed that the difference between training and validation accuracy reduces with the progression of epochs indicating proper training in the absence of both bias and variance. Furthermore, the consistent decrease in the training and validation loss indicates that the model is learning at every epoch and once the model is trained, both the curves remain unchanged. Fig. \ref{Fig: resnet50 confusion} and Fig. \ref{Fig: resnet50 roc} show the confusion matrices and the ROC curves for all the five folds. Both these figures show accurate prediction of patches with and without liver cancer. Table \ref{tab:Table3} and Table \ref{tab:Table4} show the performance metrics using the base model and the proposed hybrid model, respectively. It has been observed that all the models with the proposed hybrid strategy have resulted in better prediction of cancer grading compared to the base model. The hybrid ResNet50 model provided an increment of 1.74\% (100\% vs. 98.26\%) over the base ResNet50 model. Similarly, the other performance metrics such as sensitivity, specificity, and F1-score also provided the prediction performance of 100\% with an AUC of unity. The results indicate that the hybrid learning strategy can avoid false predictions justifying the robustness of the model. 

\begin{figure}[ht!]
\centering
\scalebox{1}{\includegraphics[width=3.4in, 
		]{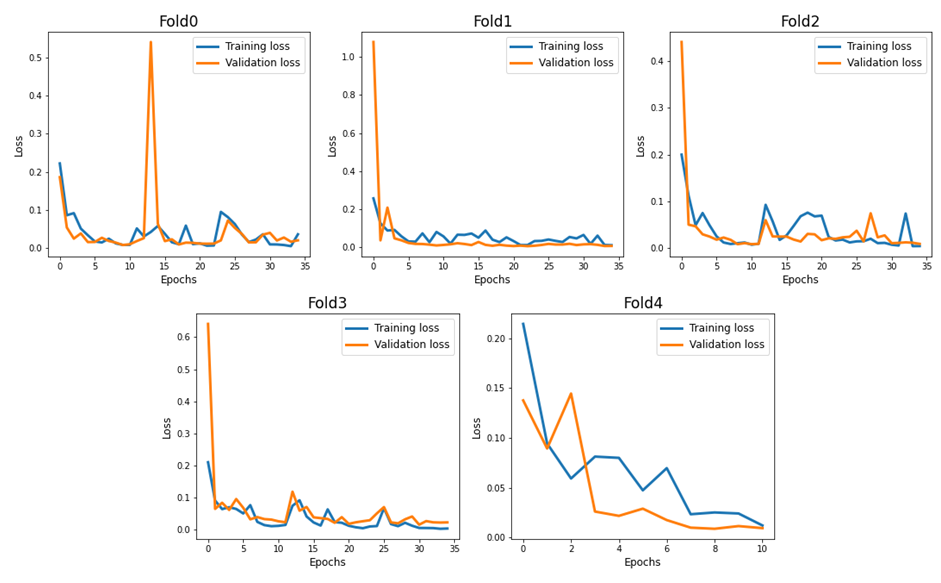}}
\caption{Train and validation loss plotted for all epochs, all 5 folds over TCGA train dataset of hybrid model with ResNet50 as feature extractor}
\label{Fig: resnet50 loss}
\end{figure}

\begin{figure}[ht!]
\centering
\scalebox{1}{\includegraphics[width=3.4in, 
		]{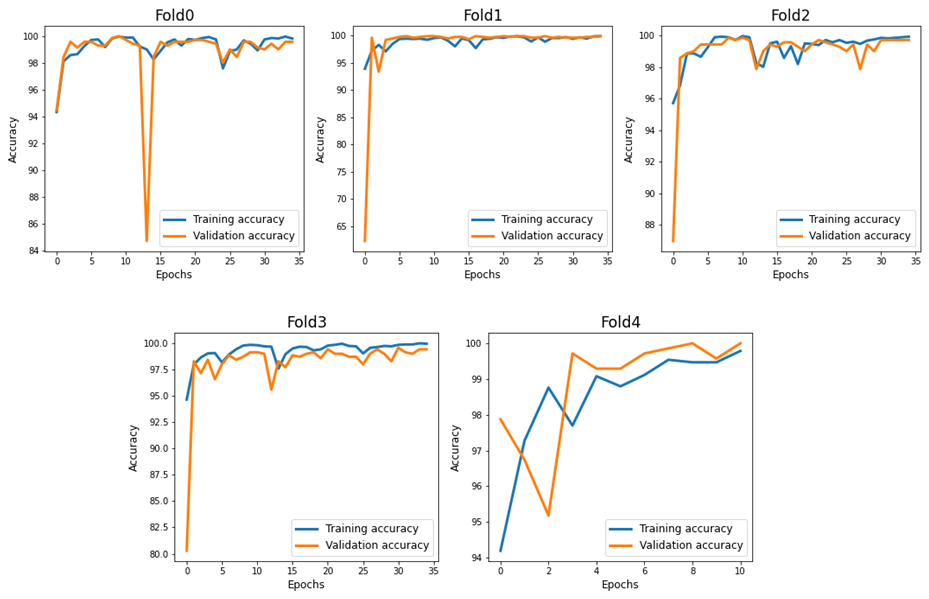}}
\caption{Train and validation accuracy plotted for all epochs, all 5 folds over TCGA train dataset of hybrid model with ResNet50 as feature extractor}
\label{Fig: resnet50 accuracy}
\end{figure}

\begin{figure}[ht!]
\centering
\scalebox{1}{\includegraphics[width=3.4in, 
		]{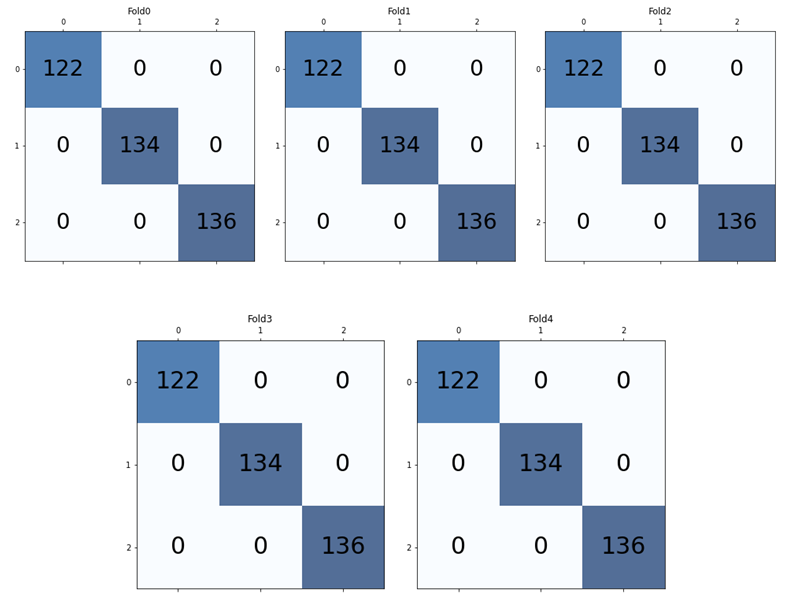}}
\caption{Confusion matrix of all 5 folds over TCGA test dataset of trained hybrid model with ResNet50 as feature extractor}
\label{Fig: resnet50 confusion}
\end{figure}

\begin{figure}[ht!]
\centering
\scalebox{1}{\includegraphics[width=3.4in, height=3.3in
		]{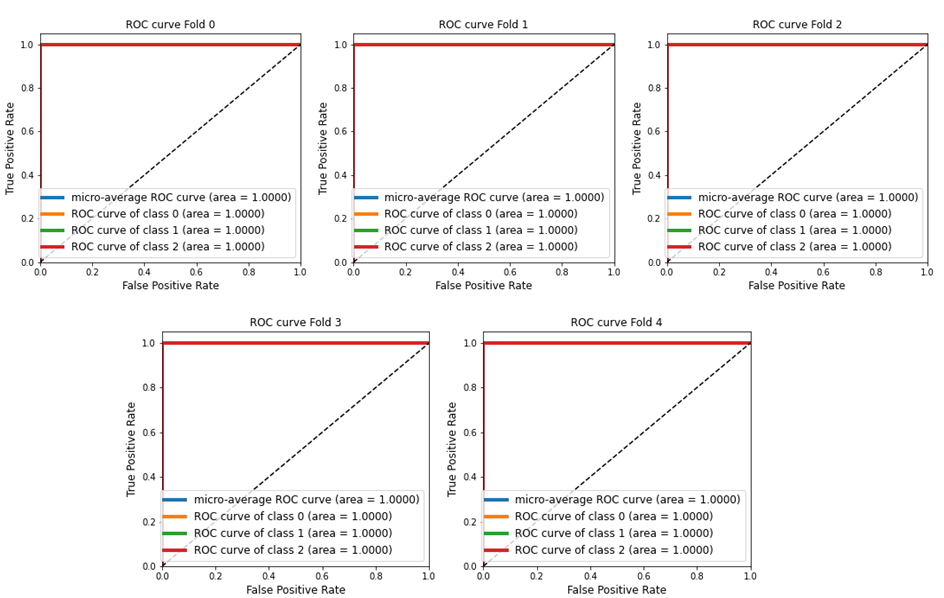}}
\caption{ROC curve of all 5 folds over TCGA test dataset of trained hybrid model with ResNet50 as a feature extractor.}
\label{Fig: resnet50 roc}
\end{figure}

\begin{table*}[ht!]
\scriptsize
\centering
\caption{The performance metrics of base models calculated on TCGA-LIHC test dataset.}
\label{tab:Table3}
\begin{adjustbox}{width=1\textwidth}
    \begin{tabular}{ l l c c c c c c c c }
    \hline
        \textbf{Metrics} & ~ & \textbf{ResNet50} & \textbf{VGG16} & \textbf{EfficientNetb0} & \textbf{EfficientNetb1} & \textbf{EfficientNetb2} & \textbf{EfficientNetb3} & \textbf{EfficientNetb4} & \textbf{DenseNet121} \\ \hline
\textbf{Accuracy}    &              & 98.26$\pm$0.11        & 98.51$\pm$0.48                      & \textbf{98.92$\pm$0.49}                      & 97.19$\pm$0.59                               & 98.41$\pm$0.37                               & 94.48$\pm$0.49                               & 95.35$\pm$0.75                               & 97.49$\pm$0.21                            \\
\textbf{F1-score}    & Type0        & 99.59$\pm$0.00        & 99.91$\pm$0.18                      & \textbf{99.59$\pm$0.00}                      & 99.59$\pm$0.40                               & 100.00$\pm$0.00                              & 99.09$\pm$0.18                               & 99.42$\pm$0.37                               & 100.00$\pm$0.00                           \\
                     & Type1        & 97.51$\pm$0.16        & 97.86$\pm$0.69                      & \textbf{98.44$\pm$0.69}                      & 95.84$\pm$0.89                               & 97.66$\pm$0.57                               & 92.33$\pm$0.88                               & 93.69$\pm$0.94                               & 96.43$\pm$0.29                            \\
                     & Type2        & 97.82$\pm$0.17        & 97.91$\pm$0.62                      & \textbf{98.81$\pm$0.71}                      & 96.36$\pm$0.74                               & 97.74$\pm$0.53                               & 92.50$\pm$0.57                               & 93.36$\pm$0.98                               & 96.30$\pm$0.32                            \\
                     & Weighted avg & 98.27$\pm$0.11        & 98.51$\pm$0.48                      & \textbf{98.93$\pm$0.48}                      & 97.19$\pm$0.59                               & 98.41$\pm$0.37                               & 94.49$\pm$0.49                               & 95.36$\pm$0.74                               & 97.49$\pm$0.21                            \\
\textbf{Specificity} & Type0        & 100.00$\pm$0.00       & 100.00$\pm$0.00                     & \textbf{100.00$\pm$0.00}                     & 99.77$\pm$0.20                               & 100.00$\pm$0.00                              & 99.70$\pm$0.16                               & 100.00$\pm$0.00                              & 100.00$\pm$0.00                           \\
                     & Type1        & 97.51$\pm$0.20        & 98.14$\pm$0.63                      & \textbf{98.75$\pm$0.74}                      & 98.45$\pm$0.47                               & 99.22$\pm$0.00                               & 95.34$\pm$0.61                               & 94.49$\pm$1.35                               & 96.74$\pm$0.34                            \\
                     & Type2        & 99.84$\pm$0.21        & 99.61$\pm$0.27                      & \textbf{99.61$\pm$0.00}                      & 97.50$\pm$0.76                               & 98.36$\pm$0.57                               & 96.56$\pm$0.64                               & 98.44$\pm$0.55                               & 99.45$\pm$0.21                            \\
                     & Macro avg    & 99.12$\pm$0.05        & 99.25$\pm$0.24                      & \textbf{99.45$\pm$0.24}                      & 98.57$\pm$0.30                               & 99.19$\pm$0.19                               & 97.20$\pm$0.25                               & 97.64$\pm$0.38                               & 98.73$\pm$0.10                            \\
\textbf{Sensitivity} & Type0        & 99.18$\pm$0.00        & 99.83$\pm$0.36                      & \textbf{99.18$\pm$0.00}                      & 99.67$\pm$0.44                               & 100.00$\pm$0.00                              & 98.85$\pm$0.44                               & 98.85$\pm$0.73                               & 100.00$\pm$0.00                           \\
                     & Type1        & 99.70$\pm$0.41        & 99.25$\pm$0.52                      & \textbf{99.25$\pm$0.00}                      & 94.77$\pm$1.39                               & 96.86$\pm$1.10                               & 93.43$\pm$1.22                               & 97.46$\pm$0.85                               & 98.95$\pm$0.40                            \\
                     & Type2        & 96.02$\pm$0.39        & 96.61$\pm$0.98                      & \textbf{98.38$\pm$1.41}                      & 97.35$\pm$0.65                               & 98.53$\pm$0.00                               & 91.62$\pm$0.65                               & 90.14$\pm$2.11                               & 93.82$\pm$0.65                            \\
                     & Macro avg    & 98.30$\pm$0.11        & 98.56$\pm$0.48                      & \textbf{98.93$\pm$0.47}                      & 97.26$\pm$0.58                               & 98.46$\pm$0.36                               & 94.63$\pm$0.49                               & 95.48$\pm$0.74                               & 97.59$\pm$0.20                            \\
\textbf{AUC}         & Type0        & 1.00$\pm$0.00         & 1.00$\pm$0.00                       & \textbf{1.00$\pm$0.00}                       & 0.99$\pm$0.00                                & 1.00$\pm$0.00                                & 0.99$\pm$0.00                                & 1.00$\pm$0.00                                & 1.00$\pm$0.00                             \\
                     & Type1        & 0.99$\pm$0.00         & 0.99$\pm$0.00                       & \textbf{0.99$\pm$0.00}                       & 0.99$\pm$0.00                                & 0.99$\pm$0.00                                & 0.98$\pm$0.00                                & 0.99$\pm$0.00                                & 0.99$\pm$0.00                             \\
                     & Type2        & 0.99$\pm$0.00         & 0.99$\pm$0.00                       & \textbf{0.99$\pm$0.00}                       & 0.99$\pm$0.00                                & 0.99$\pm$0.00                                & 0.99$\pm$0.00                                & 0.99$\pm$0.00                                & 0.99$\pm$0.00                             \\
                     & Macro avg    & 0.99$\pm$0.00         & 0.99$\pm$0.00                       & \textbf{0.99$\pm$0.00}                       & 0.99$\pm$0.00                                & 0.99$\pm$0.00                                & 0.99$\pm$0.00                                & 0.99$\pm$0.00                                & 0.99$\pm$0.00                             \\          \hline
    \multicolumn{10}{l}{*All metrics except AUC are expressed in percentages.}
    \end{tabular}
    \end{adjustbox}
\end{table*}

\begin{table*}[ht!]
\scriptsize
\centering
\caption{The performance metrics of hybrid models calculated on TCGA-LIHC test dataset.}
\label{tab:Table4}
\begin{adjustbox}{width=1\textwidth}
    \begin{tabular}{ l l c c c c c c c c }
    \hline
        \textbf{Metrics} & ~ & \textbf{ResNet50} & \textbf{VGG16} & \textbf{EfficientNetb0} & \textbf{EfficientNetb1} & \textbf{EfficientNetb2} & \textbf{EfficientNetb3} & \textbf{EfficientNetb4} & \textbf{DenseNet121} \\ \hline
        Accuracy & ~ & \textbf{100.00$\pm$0.00} & 99.43$\pm$0.49 & 99.89$\pm$0.14 & 99.94$\pm$0.11 & 99.89$\pm$0.22 & 99.74$\pm$0.18 & 99.74$\pm$0.18 & 99.59$\pm$0.53 \\ 
        f1-score & class 0 & \textbf{100.00$\pm$0.00} & 99.83$\pm$0.37 & 100.00$\pm$0.00 & 100.00$\pm$0.00 & 100.00$\pm$0.00 & 99.91$\pm$0.18 & 100.00$\pm$0.00 & 100.00$\pm$0.00 \\ 
        ~ & class 1 & \textbf{100.00$\pm$0.00} & 99.18$\pm$0.71 & 99.85$\pm$0.20 & 99.92$\pm$0.16 & 99.85$\pm$0.33 & 99.63$\pm$0.26 & 99.62$\pm$0.26 & 99.40$\pm$0.77 \\ 
        ~ & class 2 & \textbf{100.00$\pm$0.00} & 99.33$\pm$0.48 & 99.85$\pm$0.20 & 99.92$\pm$0.16 & 99.85$\pm$0.33 & 99.70$\pm$0.30 & 99.63$\pm$0.26 & 99.40$\pm$0.77 \\ 
        ~ & macro avg & \textbf{100.00$\pm$0.00} & 99.45$\pm$0.48 & 99.90$\pm$0.13 & 99.95$\pm$0.11 & 99.90$\pm$0.21 & 99.75$\pm$0.17 & 99.75$\pm$0.17 & 99.60$\pm$0.51 \\ 
        ~ & weighted avg & \textbf{100.00$\pm$0.00} & 99.43$\pm$0.48 & 99.89$\pm$0.14 & 99.94$\pm$0.11 & 99.89$\pm$0.22 & 99.74$\pm$0.18 & 99.74$\pm$0.18 & 99.59$\pm$0.53 \\ 
        Specificity & class 0 & \textbf{100.00$\pm$0.00} & 100.00$\pm$0.00 & 100.00$\pm$0.00 & 100.00$\pm$0.00 & 100.00$\pm$0.00 & 99.92$\pm$0.16 & 100.00$\pm$0.00 & 100.00$\pm$0.00 \\ 
        ~ & class 1 & \textbf{100.00$\pm$0.00} & 99.45$\pm$0.64 & 99.92$\pm$0.17 & 100.00$\pm$0.00 & 99.92$\pm$0.17 & 99.68$\pm$0.32 & 99.84$\pm$0.21 & 99.53$\pm$0.69 \\ 
        ~ & class 2 & \textbf{100.00$\pm$0.00} & 99.68$\pm$0.17 & 99.92$\pm$0.17 & 99.92$\pm$0.17 & 99.92$\pm$0.17 & 100.00$\pm$0.00 & 99.76$\pm$0.21 & 99.84$\pm$0.21 \\ 
        ~ & macro avg & \textbf{100.00$\pm$0.00} & 99.71$\pm$0.24 & 99.94$\pm$0.07 & 99.97$\pm$0.05 & 99.94$\pm$0.11 & 99.87$\pm$0.09 & 99.87$\pm$0.09 & 99.79$\pm$0.27 \\ 
        Sensitivity & class 0 & \textbf{100.00$\pm$0.00} & 99.67$\pm$0.73 & 100.00$\pm$0.00 & 100.00$\pm$0.00 & 100.00$\pm$0.00 & 100.00$\pm$0.00 & 100.00$\pm$0.00 & 100.00$\pm$0.00 \\ 
        ~ & class 1 & \textbf{100.00$\pm$0.00} & 99.40$\pm$0.33 & 99.85$\pm$0.33 & 99.85$\pm$0.33 & 99.85$\pm$0.33 & 99.85$\pm$0.33 & 99.55$\pm$0.41 & 99.70$\pm$0.41 \\ 
        ~ & class 2 & \textbf{100.00$\pm$0.00} & 99.26$\pm$0.73 & 99.85$\pm$0.33 & 100.00$\pm$0.00 & 99.85$\pm$0.33 & 99.41$\pm$0.61 & 99.70$\pm$0.40 & 99.11$\pm$1.31 \\ 
        ~ & macro avg & \textbf{100.00$\pm$0.00} & 99.44$\pm$0.49 & 99.90$\pm$0.13 & 99.95$\pm$0.11 & 99.90$\pm$0.22 & 99.75$\pm$0.17 & 99.75$\pm$0.17 & 99.60$\pm$0.51 \\ 
        AUC & class 0 & \textbf{1.00$\pm$0.00} & 1.00$\pm$0.00 & 1.00$\pm$0.00 & 1.00$\pm$0.00 & 1.00$\pm$0.00 & 1.00$\pm$0.00 & 1.00$\pm$0.00 & 1.00$\pm$0.00 \\ 
        ~ & class 1 & \textbf{1.00$\pm$0.00} & 0.99$\pm$0.00 & 1.00$\pm$0.00 & 1.00$\pm$0.00 & 1.00$\pm$0.00 & 1.00$\pm$0.00 & 0.99$\pm$0.00 & 0.99$\pm$0.00 \\ 
        ~ & class 2 & \textbf{1.00$\pm$0.00} & 0.99$\pm$0.00 & 1.00$\pm$0.00 & 1.00$\pm$0.00 & 1.00$\pm$0.00 & 1.00$\pm$0.00 & 0.99$\pm$0.00 & 0.99$\pm$0.00 \\ 
        ~ & macro avg & \textbf{1.00$\pm$0.00} & 0.99$\pm$0.00 & 1.00$\pm$0.00 & 1.00$\pm$0.00 & 1.00$\pm$0.00 & 0.99$\pm$0.00 & 0.99$\pm$0.00 & 0.99$\pm$0.00 \\ 
     \hline
         \multicolumn{10}{l}{*All metrics except AUC are expressed in percentages.}
    \end{tabular}
    \end{adjustbox}
\end{table*}

\subsubsection{Results on KMC dataset}
The training and validation accuracy and loss are shown in Fig. \ref{Fig: efficientnetb3 loss} and Fig. \ref{Fig: efficientnetb3 accuracy}, respectively. Here as well proper training in the absence of both bias and variance can be observed. Furthermore, the consistent decrease in the training and validation loss indicates that the model is learning at every epoch and once the model is trained, both the curves saturate. Fig. \ref{Fig: efficientnetb3 confusion} and Fig. \ref{Fig: efficientnetb3 roc} show the confusion matrices and the ROC curves for all five folds. Table \ref{tab:Table5} and Table \ref{tab:Table6} show the performance metrics using the base model and the proposed hybrid model, respectively over the KMC dataset. The results for KMC dataset also show that the proposed hybrid strategy has resulted in better prediction of cancer grading compared to the base model. The hybrid EfficientNetb3 achieves 96.71\% accuracy with a 2.22\% standard deviation. This is an increment of 4.65\% over the base EfficientNetb3 model. 

\begin{figure}[ht!]
\centering
\scalebox{1}{\includegraphics[width=3.4in, 
		]{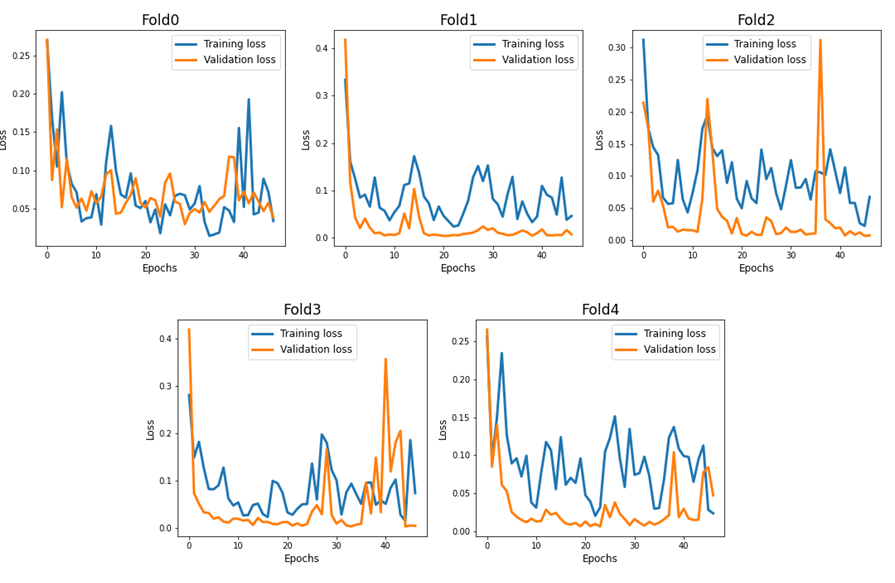}}
\caption{Training and validation loss plotted for all epochs, all 5 folds over KMC train dataset of hybrid model with EfficientNetb3 as feature extractor}
\label{Fig: efficientnetb3 loss}
\end{figure}

\begin{figure}[ht!]
\centering
\scalebox{1}{\includegraphics[width=3.4in, 
		]{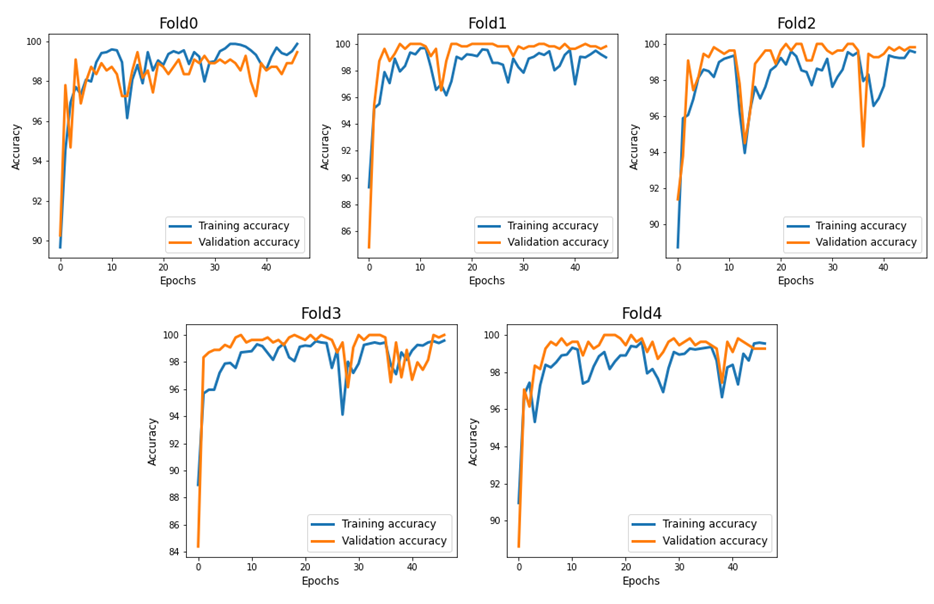}}
\caption{Training and validation accuracy plotted for all epochs, all 5 folds over KMC train dataset of hybrid model with EfficientNetb3 as feature extractor}
\label{Fig: efficientnetb3 accuracy}
\end{figure}

\begin{figure}[ht!]
\centering
\scalebox{1}{\includegraphics[width=3.4in, 
		]{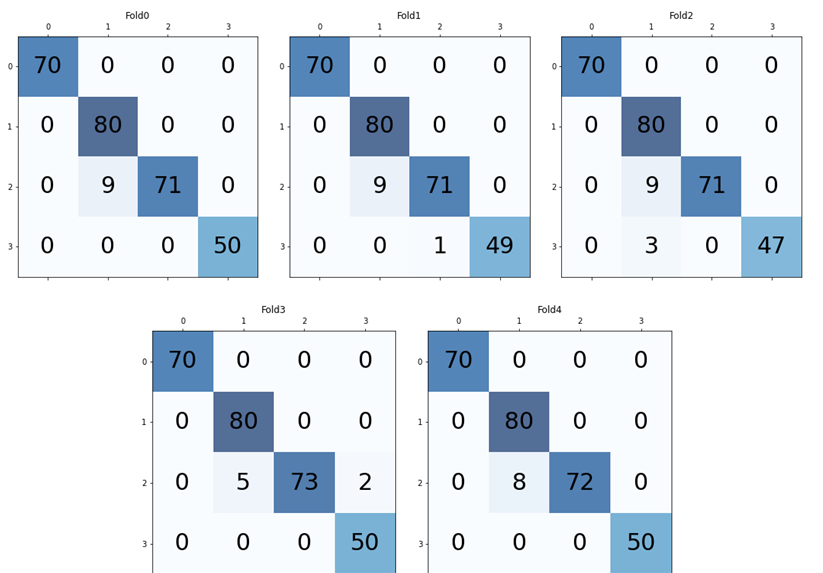}}
\caption{Confusion matrix of all 5 folds over KMC test dataset of trained hybrid model with EfficientNetb3 as feature extractor}
\label{Fig: efficientnetb3 confusion}
\end{figure}

\begin{figure}[ht!]
\centering
\scalebox{1}{\includegraphics[width=3.4in, 
		]{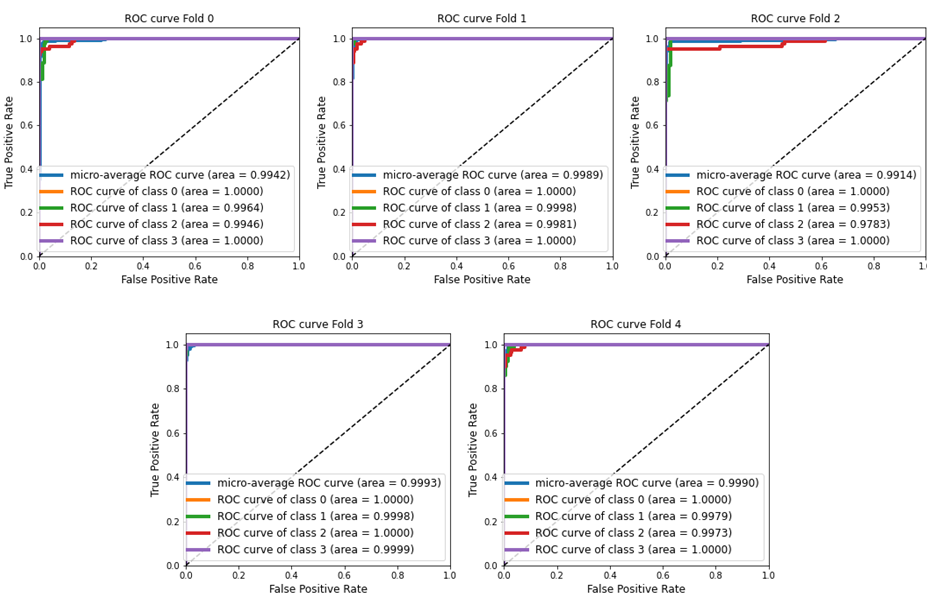}}
\caption{ROC curve of all 5 folds over KMC test dataset of trained hybrid model with EfficientNetb3 as feature extractor.}
\label{Fig: efficientnetb3 roc}
\end{figure}

\begin{table*}[ht!]
\scriptsize
\centering
\caption{The performance metrics of base models calculated on KMC test dataset. }
\label{tab:Table5}
\begin{adjustbox}{width=1\textwidth}
    \begin{tabular}{ l l c c c c c c c c }
    \hline
        \textbf{Metrics} & ~ & \textbf{ResNet50} & \textbf{VGG16} & \textbf{EfficientNetb0} & \textbf{EfficientNetb1} & \textbf{EfficientNetb2} & \textbf{EfficientNetb3} & \textbf{EfficientNetb4} & \textbf{DenseNet121} \\ \hline
Accuracy   & ~  & 88.28$\pm$0.63  & 87.35$\pm$0.59   & 91.49$\pm$0.73    & 92.57$\pm$0.77   & 88.28$\pm$1.32  & \textbf{92.07$\pm$0.81}  & 90.42$\pm$0.68 & 89.78$\pm$0.85 \\
F1-score    & Type0 & 100.00$\pm$0.00   & 100.00$\pm$0.00 & 99.29$\pm$0.00  & 99.85$\pm$0.31  & 99.14$\pm$0.32  & \textbf{99.00$\pm$0.38}  & 98.85$\pm$1.08  & 100.00$\pm$0.00 \\
& Type1 & 81.70$\pm$0.83 & 83.75$\pm$1.49 & 86.78$\pm$1.18  & 88.51$\pm$0.88  & 81.18$\pm$2.06 & \textbf{88.23$\pm$0.64} & 87.07$\pm$1.13 & 84.52$\pm$1.62 \\
& Type2 & 84.87$\pm$1.36 & 78.65$\pm$1.12 & 86.55$\pm$1.05 & 87.18$\pm$1.44  & 79.78$\pm$1.82  & \textbf{86.91$\pm$1.25} & 86.09$\pm$1.04  & 87.19$\pm$1.16   \\
 & Type3 & 88.89$\pm$0.00   & 89.37$\pm$0.66    & 96.26$\pm$0.96   & 97.53$\pm$0.95  & 98.16$\pm$0.86 & \textbf{96.89$\pm$1.31}  & 91.30$\pm$0.00  & 88.64$\pm$0.55 \\
 & Weighted  avg & 88.46$\pm$0.62  & 87.36$\pm$0.59  & 91.53$\pm$0.72  & 92.57$\pm$0.77  & 88.30$\pm$1.27  & \textbf{92.09$\pm$0.81}  & 90.49$\pm$0.68   & 89.89$\pm$0.82  \\
Specificity & Type0  & 100.00$\pm$0.00  & 100.00$\pm$0.00 & 99.52$\pm$0.00 & 99.90$\pm$0.21 & 99.52$\pm$0.00 & \textbf{99.42$\pm$0.21} & 99.61$\pm$0.39 & 100.00$\pm$0.00  \\
& Type1 & 87.00$\pm$0.86  & 87.50$\pm$2.29  & 93.70$\pm$0.27 & 91.50$\pm$0.50  & 91.30$\pm$0.57 & \textbf{94.90$\pm$0.82} & 94.10$\pm$0.41  & 88.90$\pm$1.14  \\
 & Type2   & 96.60$\pm$0.22  & 94.80$\pm$2.01 & 94.90$\pm$1.08  & 98.20$\pm$0.75 & 92.80$\pm$2.07 & \textbf{94.60$\pm$0.54}  & 92.90$\pm$0.82 & 96.80$\pm$0.44 \\
 & Type3   & 100.00$\pm$0.00   & 100.00$\pm$0.00  & 100.00$\pm$0.00  & 100.00$\pm$0.00 & 100.00$\pm$0.00 & \textbf{100.00$\pm$0.00} & 100.00$\pm$0.00   & 100.00$\pm$0.00  \\
 & Macro avg    & 95.90$\pm$0.22  & 95.57$\pm$0.20  & 97.03$\pm$0.25   & 97.40$\pm$0.27  & 95.90$\pm$0.46  & \textbf{97.23$\pm$0.28} & 96.65$\pm$0.23 & 96.42$\pm$0.30 \\
Sensitivity                                & Type0        & 100.00$\pm$0.00                                 & 100.00$\pm$0.00                     & 100.00$\pm$0.00                              & 100.00$\pm$0.00                              & 99.71$\pm$0.63                               & \textbf{99.71$\pm$0.63}                      & 98.85$\pm$1.19                               & 100.00$\pm$0.00                           \\
                                           & Type1        & 91.50$\pm$0.55                                  & 94.50$\pm$2.59                      & 88.75$\pm$2.33                               & 96.25$\pm$0.88                               & 83.25$\pm$4.10                               & \textbf{89.00$\pm$1.04}                      & 88.50$\pm$2.40                               & 93.50$\pm$1.62                            \\
                                           & Type2        & 80.00$\pm$2.16                                  & 73.25$\pm$3.37                      & 86.00$\pm$1.04                               & 80.75$\pm$1.11                               & 78.25$\pm$1.42                               & \textbf{87.25$\pm$2.05}                      & 89.00$\pm$1.04                               & 83.50$\pm$1.85                            \\
                                           & Type3        & 80.00$\pm$0.00                                  & 80.80$\pm$1.09                      & 92.80$\pm$1.78                               & 95.20$\pm$1.78                               & 96.40$\pm$1.67                               & \textbf{94.00$\pm$2.44}                      & 84.00$\pm$0.00                               & 79.60$\pm$0.89                            \\
                                           & Macro avg    & 87.87$\pm$0.55                                  & 87.13$\pm$0.52                      & 91.88$\pm$0.79                               & 93.05$\pm$0.82                               & 89.40$\pm$1.30                               & \textbf{92.49$\pm$0.92}                      & 90.08$\pm$0.61                               & 89.15$\pm$0.82                            \\
AUC                                        & Type0        & 1.00$\pm$0.00                                   & 1.00$\pm$0.00                       & 1.00$\pm$0.00                                & 1.00$\pm$0.00                                & 0.99$\pm$0.00                                & \textbf{1.00$\pm$0.00}                       & 0.99$\pm$0.00                                & 1.00$\pm$0.00                             \\
                                           & Type1        & 0.96$\pm$0.00                                   & 0.96$\pm$0.00                       & 0.98$\pm$0.00                                & 0.98$\pm$0.00                                & 0.96$\pm$0.00                                & \textbf{0.98$\pm$0.00}                       & 0.98$\pm$0.00                                & 0.97$\pm$0.00                             \\
                                           & Type2        & 0.97$\pm$0.00                                   & 0.95$\pm$0.00                       & 0.98$\pm$0.00                                & 0.97$\pm$0.00                                & 0.95$\pm$0.00                                & \textbf{0.97$\pm$0.00}                       & 0.97$\pm$0.00                                & 0.98$\pm$0.00                             \\
                                           & Type3        & 0.98$\pm$0.00                                   & 0.99$\pm$0.00                       & 0.99$\pm$0.00                                & 1.00$\pm$0.00                                & 0.99$\pm$0.00                                & \textbf{0.99$\pm$0.00}                       & 0.99$\pm$0.00                                & 0.99$\pm$0.00                             \\
                                           & Macro avg    & 0.98$\pm$0.00                                   & 0.98$\pm$0.00                       & 0.99$\pm$0.00                                & 0.99$\pm$0.00                                & 0.98$\pm$0.00                                & \textbf{0.99$\pm$0.00}                       & 0.99$\pm$0.00       & 0.98$\pm$0.00 \\  \hline
    \multicolumn{10}{l}{*All metrics except AUC are expressed in percentages.}
    \end{tabular}
    \end{adjustbox}
\end{table*}

\begin{table*}[ht!]
\scriptsize
\centering
\caption{The performance metrics of hybrid models calculated on KMC test dataset. }
\label{tab:Table6}
\begin{adjustbox}{width=1\textwidth}
    \begin{tabular}{ l l c c c c c c c c }
    \hline
        \textbf{Metrics} & ~ & \textbf{ResNet50} & \textbf{VGG16} & \textbf{EfficientNetb0} & \textbf{EfficientNetb1} & \textbf{EfficientNetb2} & \textbf{EfficientNetb3} & \textbf{EfficientNetb4} & \textbf{DenseNet121} \\ \hline
        \textbf{Accuracy} & ~ & 95.07$\pm$2.13 & 90.64$\pm$1.24 & 96.21$\pm$1.81 & 96.14$\pm$2.24 & 95.49$\pm$2.22 & \textbf{96.71$\pm$0.68} & 96.49$\pm$2.41 & 94.92$\pm$1.14 \\ 
        \textbf{F1-score} & Type0 & 98.79$\pm$2.70 & 99.29$\pm$0.70 & 99.58$\pm$0.93 & 100.00$\pm$0.00 & 100.00$\pm$0.00 & \textbf{100.00$\pm$0.00} & 100.00$\pm$0.00 & 99.58$\pm$0.94 \\ 
        ~ & Type1 & 93.01$\pm$1.69 & 88.74$\pm$2.41 & 94.48$\pm$1.83 & 94.76$\pm$3.12 & 93.40$\pm$2.75 & \textbf{94.91$\pm$1.41} & 97.70$\pm$1.66 & 94.26$\pm$2.12 \\ 
        ~ & Type2 & 93.35$\pm$2.43 & 85.95$\pm$3.84 & 92.90$\pm$3.51 & 94.02$\pm$2.99 & 93.65$\pm$3.98 & \textbf{94.33$\pm$0.76} & 94.25$\pm$3.93 & 92.40$\pm$3.02 \\ 
        ~ & Type3 & 95.80$\pm$5.76 & 88.75$\pm$1.83 & 99.38$\pm$1.38 & 96.38$\pm$3.51 & 95.42$\pm$5.09 & \textbf{98.78$\pm$1.32} & 92.96$\pm$5.06 & 93.24$\pm$4.39 \\ 
        ~ & Weighted avg & 95.05$\pm$2.17 & 90.58$\pm$1.24 & 96.17$\pm$1.85 & 96.15$\pm$2.22 & 95.48$\pm$2.23 & \textbf{96.71$\pm$0.66} & 96.44$\pm$2.46 & 94.88$\pm$1.16 \\ 
        \textbf{Specificity} & Type0 & 99.14$\pm$1.91 & 99.52$\pm$0.47 & 99.71$\pm$0.63 & 100.00$\pm$0.00 & 100.00$\pm$0.00 & \textbf{100.00$\pm$0.00} & 100.00$\pm$0.00 & 99.71$\pm$0.64 \\ 
        ~ & Type1 & 95.00$\pm$2.00 & 91.60$\pm$1.29 & 95.30$\pm$1.64 & 95.50$\pm$2.85 & 94.40$\pm$2.67 & \textbf{95.70$\pm$1.25} & 98.10$\pm$1.38 & 95.10$\pm$1.92 \\ 
        ~ & Type2 & 99.00$\pm$1.17 & 96.50$\pm$1.36 & 99.70$\pm$0.67 & 99.10$\pm$1.24 & 99.30$\pm$0.57 & \textbf{99.90$\pm$0.22} & 97.00$\pm$2.15 & 98.10$\pm$2.07 \\ 
        ~ & Type3 & 100.00$\pm$0.00 & 99.39$\pm$0.72 & 100.00$\pm$0.00 & 100.00$\pm$0.00 & 100.00$\pm$0.00 & \textbf{99.82$\pm$0.38} & 100.00$\pm$0.00 & 100.00$\pm$0.00 \\ 
        ~ & Macro avg & 98.28$\pm$0.72 & 96.75$\pm$0.43 & 98.67$\pm$0.62 & 98.65$\pm$0.78 & 98.42$\pm$0.77 & \textbf{98.85$\pm$0.24} & 98.77$\pm$0.84 & 98.22$\pm$0.40 \\ 
        \textbf{Sensitivity} & Type0 & 100.00$\pm$0.00 & 100.00$\pm$0.00 & 100.00$\pm$0.00 & 100.00$\pm$0.00 & 100.00$\pm$0.00 & \textbf{100.00$\pm$0.00} & 100.00$\pm$0.00 & 100.00$\pm$0.00 \\ 
        ~ & Type1 & 97.75$\pm$2.56 & 96.50$\pm$2.40 & 100.00$\pm$0.00 & 100.00$\pm$0.00 & 99.75$\pm$0.55 & \textbf{100.00$\pm$0.00} & 100.00$\pm$0.00 & 100.00$\pm$0.00 \\ 
        ~ & Type2 & 89.75$\pm$3.89 & 82.00$\pm$4.47 & 87.50$\pm$5.07 & 90.75$\pm$3.81 & 89.75$\pm$5.95 & \textbf{89.50$\pm$1.11} & 95.75$\pm$2.87 & 90.00$\pm$5.45 \\ 
        ~ & Type3 & 92.40$\pm$10.43 & 82.00$\pm$2.00 & 98.80$\pm$2.68 & 93.20$\pm$6.57 & 91.60$\pm$9.20 & \textbf{98.40$\pm$2.60} & 87.20$\pm$9.01 & 87.60$\pm$7.92 \\ 
        ~ & Macro avg & 94.97$\pm$2.73 & 90.12$\pm$0.91 & 96.57$\pm$1.79 & 95.98$\pm$2.58 & 95.27$\pm$2.61 & \textbf{96.97$\pm$0.83} & 95.73$\pm$2.95 & 94.40$\pm$1.40 \\ 
        \textbf{AUC} & Type0 & 1.00$\pm$0.00 & 1.00$\pm$0.00 & 1.00$\pm$0.00 & 1.00$\pm$0.00 & 1.00$\pm$0.00 & \textbf{1.00$\pm$0.00} & 1.00$\pm$0.00 & 1.00$\pm$0.00 \\ 
        ~ & Type1 & 0.98$\pm$0.01 & 0.98$\pm$0.00 & 0.99$\pm$0.00 & 0.99$\pm$0.00 & 0.99$\pm$0.00 & \textbf{0.99$\pm$0.00} & 0.99$\pm$0.00 & 0.99$\pm$0.00 \\ 
        ~ & Type2 & 0.98$\pm$0.02 & 0.97$\pm$0.00 & 0.99$\pm$0.00 & 0.99$\pm$0.00 & 0.99$\pm$0.00 & \textbf{0.99$\pm$0.00} & 0.99$\pm$0.00 & 0.99$\pm$0.00 \\ 
        ~ & Type3 & 0.99$\pm$0.00 & 0.98$\pm$0.01 & 1.00$\pm$0.00 & 1.00$\pm$0.00 & 0.99$\pm$0.00 & \textbf{0.99$\pm$0.00} & 0.99$\pm$0.00 & 0.99$\pm$0.00 \\ 
        ~ & Macro avg & 0.99$\pm$0.00 & 0.98$\pm$0.00 & 0.99$\pm$0.00 & 0.99$\pm$0.00 & 0.99$\pm$0.00 & \textbf{0.99$\pm$0.00} & 0.99$\pm$0.00 & 0.99$\pm$0.00 \\ 
    \hline
    \multicolumn{10}{l}{*All metrics except AUC are expressed in percentages.}
    \end{tabular}
    \end{adjustbox}
\end{table*}

\subsubsection{Results on COLON dataset}
To assess the performance of the proposed hybrid model, colon dataset is also considered. Fig. \ref{Fig: denseset121 confusion} and Fig. \ref{Fig: densenet121 roc} show the confusion matrices and the ROC curves for all five folds. Table \ref{tab:Table7} and Table \ref{tab:Table8} show the performance metrics using the base model and the proposed hybrid model, respectively over the COLON dataset. Hybrid models are found to perform better for this database as well. The hybrid  EfficientNetb2, EfficientNetb4 and DenseNet121 achieve 100\% accuracy. This is an increment of 0.44, 0.36, and 0.32\% over the base models respectively. 

\begin{figure}[ht!]
\centering
\scalebox{1}{\includegraphics[width=2.8in, 
		]{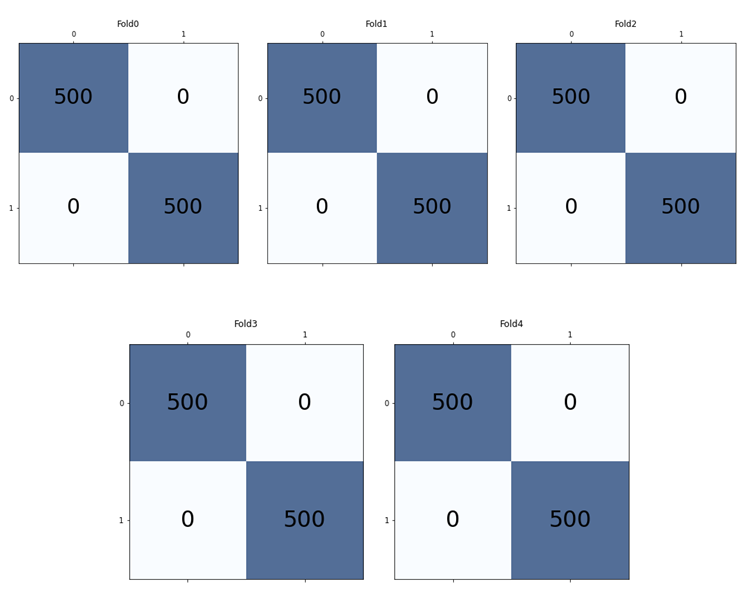}}
\caption{Confusion matrix of all 5 folds over COLON test dataset of trained hybrid model with DenseNet121 as feature extractor}
\label{Fig: denseset121 confusion}
\end{figure}

\begin{figure}[ht!]
\centering
\scalebox{1}{\includegraphics[width=3.4in, 
		]{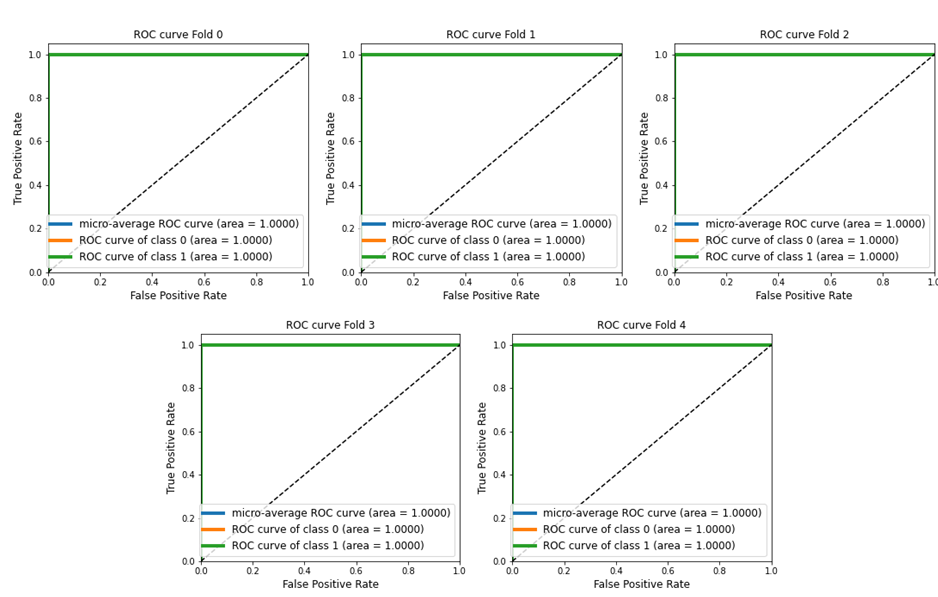}}
\caption{ROC curve of all 5 folds over COLON test dataset of trained hybrid model with DenseNet121 as feature extractor}
\label{Fig: densenet121 roc}
\end{figure}

\begin{table*}[ht!]
\scriptsize
\centering
\caption{The performance metrics of base models calculated on COLON test dataset.}
\label{tab:Table7}
\begin{adjustbox}{width=1\textwidth}
    \begin{tabular}{ l l c c c c c c c c }
    \hline
        \textbf{Metrics} & ~ & \textbf{ResNet50} & \textbf{VGG16} & \textbf{EfficientNetb0} & \textbf{EfficientNetb1} & \textbf{EfficientNetb2} & \textbf{EfficientNetb3} & \textbf{EfficientNetb4} & \textbf{DenseNet121} \\ \hline
Accuracy                                   &              & 99.54$\pm$0.13                                  & \textbf{99.94$\pm$0.05}             & 99.90$\pm$0.07                               & 99.70$\pm$0.07                               & 99.56$\pm$0.18                               & 99.84$\pm$0.05                               & 99.74$\pm$0.09                               & 99.68$\pm$0.18                            \\
F1-score                                   & Type0        & 99.54$\pm$0.13                                  & \textbf{99.94$\pm$0.05}             & 99.90$\pm$0.07                               & 99.70$\pm$0.07                               & 99.56$\pm$0.18                               & 99.84$\pm$0.05                               & 99.74$\pm$0.09                               & 99.68$\pm$0.18                            \\
                                           & Type1        & 99.54$\pm$0.13                                  & \textbf{99.94$\pm$0.05}             & 99.90$\pm$0.07                               & 99.70$\pm$0.07                               & 99.56$\pm$0.18                               & 99.84$\pm$0.05                               & 99.74$\pm$0.09                               & 99.68$\pm$0.18                            \\
                                           & Weighted avg & 99.54$\pm$0.13                                  & \textbf{99.94$\pm$0.05}             & 99.90$\pm$0.07                               & 99.70$\pm$0.07                               & 99.56$\pm$0.18                               & 99.84$\pm$0.05                               & 99.74$\pm$0.09                               & 99.68$\pm$0.18                            \\
Specificity                                & Type0        & 100.00$\pm$0.00                                 & \textbf{100.00$\pm$0.00}            & 100.00$\pm$0.00                              & 99.80$\pm$0.00                               & 99.80$\pm$0.00                               & 100.00$\pm$0.00                              & 100.00$\pm$0.00                              & 100.00$\pm$0.00                           \\
                                           & Type1        & 99.08$\pm$0.27                                  & \textbf{99.88$\pm$0.11}             & 99.80$\pm$0.14                               & 99.60$\pm$0.14                               & 99.32$\pm$0.36                               & 99.68$\pm$0.11                               & 99.48$\pm$0.18                               & 99.36$\pm$0.36                            \\
                                           & Macro avg    & 99.54$\pm$0.13                                  & \textbf{99.94$\pm$0.05}             & 99.90$\pm$0.07                               & 99.70$\pm$0.07                               & 99.56$\pm$0.18                               & 99.84$\pm$0.05                               & 99.74$\pm$0.09                               & 99.68$\pm$0.18                            \\
Sensitivity                                & Type0        & 99.08$\pm$0.27                                  & \textbf{99.88$\pm$0.11}             & 99.80$\pm$0.14                               & 99.60$\pm$0.14                               & 99.32$\pm$0.36                               & 99.68$\pm$0.11                               & 99.48$\pm$0.18                               & 99.36$\pm$0.36                            \\
                                           & Type1        & 100.00$\pm$0.00                                 & \textbf{100.00$\pm$0.00}            & 100.00$\pm$0.00                              & 99.80$\pm$0.00                               & 99.80$\pm$0.00                               & 100.00$\pm$0.00                              & 100.00$\pm$0.00                              & 100.00$\pm$0.00                           \\
                                           & Macro avg    & 99.54$\pm$0.13                                  & \textbf{99.94$\pm$0.05}             & 99.90$\pm$0.07                               & 99.70$\pm$0.07                               & 99.56$\pm$0.18                               & 99.84$\pm$0.05                               & 99.74$\pm$0.09                               & 99.68$\pm$0.18                            \\
AUC                                        & Type0        & 1.00$\pm$0.00                                   & \textbf{1.00$\pm$0.00}              & 1.00$\pm$0.00                                & 1.00$\pm$0.00                                & 1.00$\pm$0.00                                & 1.00$\pm$0.00                                & 1.00$\pm$0.00                                & 1.00$\pm$0.00                             \\
                                           & Type1        & 1.00$\pm$0.00                                   & \textbf{1.00$\pm$0.00}              & 1.00$\pm$0.00                                & 1.00$\pm$0.00                                & 1.00$\pm$0.00                                & 1.00$\pm$0.00                                & 1.00$\pm$0.00                                & 1.00$\pm$0.00                             \\
                                           & Macro avg    & 1.00$\pm$0.00                                   & \textbf{1.00$\pm$0.00}              & 1.00$\pm$0.00                                & 1.00$\pm$0.00                                & 1.00$\pm$0.00    & 1.00$\pm$0.00  & 1.00$\pm$0.00     & 1.00$\pm$0.00                             \\ \hline
    \multicolumn{10}{l}{*All metrics except AUC are expressed in percentages.}
    \end{tabular}
    \end{adjustbox}
\end{table*}

\begin{table*}[ht!]
\scriptsize
\centering
\caption{The performance metrics of hybrid models calculated on COLON test dataset. }
\label{tab:Table8}
\begin{adjustbox}{width=1\textwidth}
    \begin{tabular}{ l l c c c c c c c c }
    \hline
        \textbf{Metrics} & ~ & \textbf{ResNet50} & \textbf{VGG16} & \textbf{EfficientNetb0} & \textbf{EfficientNetb1} & \textbf{EfficientNetb2} & \textbf{EfficientNetb3} & \textbf{EfficientNetb4} & \textbf{DenseNet121} \\ \hline
\textbf{Accuracy} &  ~  & 99.96$\pm$0.05  & 99.96$\pm$0.05 & 99.98$\pm$0.04 & 99.98$\pm$0.04 & \textbf{100.00$\pm$0.00} & 99.90$\pm$0.17  & \textbf{100.00$\pm$0.00}  & \textbf{100.00$\pm$0.00}  \\
\textbf{F1-score}  & Type0  & 99.96$\pm$0.05 & 99.96$\pm$0.05 & 99.98$\pm$0.04 & 99.98$\pm$0.04  & \textbf{100.00$\pm$0.00} & 99.90$\pm$0.17 & \textbf{100.00$\pm$0.00}   & \textbf{100.00$\pm$0.00} \\
& Type1        & 99.96$\pm$0.05  & 99.96$\pm$0.05 & 99.98$\pm$0.04 & 99.98$\pm$0.04  & \textbf{100.00$\pm$0.00} & 99.90$\pm$0.17    & \textbf{100.00$\pm$0.00}  & \textbf{100.00$\pm$0.00}  \\
& Weighted avg & 99.96$\pm$0.05  & 99.96$\pm$0.05   & 99.98$\pm$0.04 & 99.98$\pm$0.04  & \textbf{100.00$\pm$0.00}  & 99.90$\pm$0.17  & \textbf{100.00$\pm$0.00}   & \textbf{100.00$\pm$0.00}  \\
\textbf{Specificity} & Type0 & 100.00$\pm$0.00 & 100.00$\pm$0.00  & 100.00$\pm$0.00 & 100.00$\pm$0.00   & \textbf{100.00$\pm$0.00}  & 100.00$\pm$0.00 & \textbf{100.00$\pm$0.00}  & \textbf{100.00$\pm$0.00} \\
& Type1 & 99.92$\pm$0.11 & 99.92$\pm$0.11     & 99.96$\pm$0.08 & 99.96$\pm$0.09 & \textbf{100.00$\pm$0.00} & 99.80$\pm$0.35 & \textbf{100.00$\pm$0.00} & \textbf{100.00$\pm$0.00} \\
 & Macro avg    & 99.96$\pm$0.05 & 99.96$\pm$0.05 & 99.98$\pm$0.04  & 99.98$\pm$0.04 & \textbf{100.00$\pm$0.00} & 99.90$\pm$0.17 & \textbf{100.00$\pm$0.00} & \textbf{100.00$\pm$0.00}  \\
\textbf{Sensitivity} & Type0  & 99.92$\pm$0.11 & 99.92$\pm$0.11  & 99.96$\pm$0.08  & 99.96$\pm$0.09 & \textbf{100.00$\pm$0.00} & 99.80$\pm$0.35 & \textbf{100.00$\pm$0.00} & \textbf{100.00$\pm$0.00} \\
 & Type1 & 100.00$\pm$0.00  & 100.00$\pm$0.00 & 100.00$\pm$0.00 & 100.00$\pm$0.00 & \textbf{100.00$\pm$0.00} & 100.00$\pm$0.00 & \textbf{100.00$\pm$0.00} & \textbf{100.00$\pm$0.00}  \\
 & Macro avg    & 99.96$\pm$0.05  & 99.96$\pm$0.05  & 99.98$\pm$0.04 & 99.98$\pm$0.04  & \textbf{100.00$\pm$0.00}  & 99.90$\pm$0.17 & \textbf{100.00$\pm$0.00} & \textbf{100.00$\pm$0.00}\\
\textbf{AUC}  & Type0   & 1.00$\pm$0.00 & 1.00$\pm$0.00 & 1.00$\pm$0.00 & 1.00$\pm$0.00 & \textbf{1.00$\pm$0.00} & 1.00$\pm$0.00 & \textbf{1.00$\pm$0.00} & \textbf{1.00$\pm$0.00} \\
 & Type1 & 1.00$\pm$0.00  & 1.00$\pm$0.00  & 1.00$\pm$0.00 & 1.00$\pm$0.00  & \textbf{1.00$\pm$0.00}  & 1.00$\pm$0.00  & \textbf{1.00$\pm$0.00} & \textbf{1.00$\pm$0.00}  \\
 & Macro avg    & 1.00$\pm$0.00  & 1.00$\pm$0.00 & 1.00$\pm$0.00 & 1.00$\pm$0.00 & \textbf{1.00$\pm$0.00} & 1.00$\pm$0.00 & \textbf{1.00$\pm$0.00} & \textbf{1.00$\pm$0.00} \\ \hline
    \multicolumn{10}{l}{*All metrics except AUC are expressed in percentages.}
    \end{tabular}
    \end{adjustbox}
\end{table*}

\section{Discussions}

The goal of this study is to develop a methodology for classifying different types of liver HCC tumors. We proposed a deep learning-based hybrid architecture, trained and tested over two liver HCC datasets. The different data pre-processing steps along with training and validation techniques are also incorporated to build a robust model. After training the models, performance metrics over the untouched test dataset were generated and observed. In this section, conclusions made from the obtained results, a comparison of the proposed methodology with recent studies and limitations of the proposed methodology are mentioned.

We observe that the hybrid model performs better than the base model for all datasets mentioned above. The hybrid model shows a minimum 0.76\% (VGG16) and 3.29\% (VGG16) of hike in accuracy on the TCGA and KMC datasets, respectively. The hybrid model with ResNet50 as a feature extractor outperforms the others on TCGA, achieving 100\% accuracy with a 1.74\% increase after modification (Tables \ref{tab:Table3}, \ref{tab:Table4}) for the TCGA dataset. Performance metrics indicate the excellent ability of the classification and the robustness of the model. 
The hybrid model with EfficientNetb3 outperforms all other models on KMC in terms of every performance metric (Tables \ref{tab:Table5}, \ref{tab:Table6}). With an accuracy of 96.71\% and a 2.22\% standard deviation, this model sets the benchmark for KMC database.
A similar trend is observed for the COLON dataset as well. The hybrid models with EfficientNetb2, EfficientNetb4 and DenseNet121 feature extractors show remarkable performance by achieving 100\% accuracy on the test dataset. These results prove that replacing a shallow classifier with a deep classifier and fine-tuning the top layers of the feature extractor is a fruitful strategy. 

For the TCGA dataset, the base model with the EfficientNetb0 as a feature extractor gives better accuracy, whereas, among the hybrid models, ResNet50 is proven to be the best. On the other hand, for the KMC dataset, EfficientNetb3 outperforms other architectures. Various architectures are observed to produce the best performance across the datasets. This concludes that, model performance is inherently dependent on dataset-specific characteristics, precluding the existence of a universally optimal architecture across diverse data distributions. Various architectures are designed in a way to identify or detect certain types of features that are associated with particular datasets, even when the tissues are those of the liver, but share different genetic patterns and different ways of WSI preparation procedures. 

Some recent studies on liver HCC detection involving the TCGA and KMC datasets are mentioned in Table \ref{tab:Table9}. As can be seen, the proposed methodology outperforms other studies on local image classification tasks in terms of accuracy as well as other performance metrics. LiverNet architecture \cite{LiverNet} claimed the best performance over KMC and TCGA datasets in the patch-level classification task. The LiverNet model has some key features such as CBAM, ASPP block and hypercolumn technique, which take care of a lesser number of model parameters and show improved performance. Our proposed methodology exceeds LiverNet's accuracy by more than 2\% over the TCGA dataset and by more than 5\% on the KMC dataset, despite having less number of patches in the KMC dataset. LiverNet achieves F1-score of 97.72\% and 90.93\%. On the other hand, our methodology achieves 100\% and 96.446\%  F1-score  on the TCGA and KMC, respectively. They had also produced results using a very similar architecture called BreastNet\cite{BreastNet}, which was specifically designed for breast cancer classification over BreaKHis dataset\cite{Spanhol2016}. Our proposed model outperforms BreastNet as well.

Chen et al.\cite{Chen2020} used pre-trained InceptionV3 which achieves low accuracy (89.6\%) despite having a large number of parameters. They trained the model using a massive dataset containing approximately $39k$ train patches. Sun et al.\cite{global} used pre-trained ResNet50 for patch-level feature extraction. They proposed a global-level (whole slide image level) classification mechanism based on the selection of $k$-top and $k$-bottom features from a sorted aggregation of patch-level features. On the other hand, our research focuses on patch-level classification. For aggregation, Sun et al. performed p-norm pooling, and for the classification of $2k$ features, they trained a multi-layered perceptron. The $k$ value was chosen experimentally. For only two-class classification at the global level, this study claimed 100\% accuracy.

\begin{table*}[!ht]
    \centering
    \label{tab:Table9}
    \caption{Comparison with recent studies on HCC liver cancer histopathology images (i) Indicated the author and year of publication in the first column. (ii) Dataset specifications. (iii) The number of cancer types into which it was classified. (iv) Model Specifications. (v) Performance metrics used in that particular study. (vi) Training method (vii) Remarks show some special characteristics of the model architecture and methodology.}
    \begin{adjustbox}{width=1\textwidth}
    \begin{tabular}{l l l l l l l} 
    \hline
        \textbf{First Author(year)} & \textbf{Dataset} & \textbf{Types} & \textbf{Model} & \textbf{Metrics} & \textbf{Training method} & \textbf{Remark} \\ \hline

        Sun et al.  & TCGA & 2 & Pretrained & Accuracy 100 & 10 fold & Transfer learning, \\
        (2019)\cite{global}  & 189,531 train, & ~ & ResNet50 & Precision 100 & CV$^{\mathrm{b}}$ & k feature selection\\
        ~ & patches  & ~ & ~ & Recall 100 & ~ &  from sorted aggregated patch \\
        ~ & 224x224 & ~ & ~ & F1-score 100 & ~ & features, global level \\
        ~ & tested over 70WSI & ~ & ~ & ~ & ~ & classification\\ \hline

        Chen et al.  & TCGA & 3 & Pretrained & Accuracy 89.6 & 0.85:0.15 train & Transfer learning, \\
        (2020)\cite{Chen2020}  & 39k train, & ~ & Inceptionv3 & Precision 87.9 & test split & EasyDL framework\\
        ~ & patches, 9k test & ~ & ~ & Recall 77.1 & ~ & ~ \\
        ~ & 256 × 256 & ~ & ~ & F1-score 82.0 & ~ & very big train set\\ 
        ~ & ~ & ~ & ~ & MCC$^{\mathrm{a}}$ 0.912 & ~ & Huge number of paramenters\\ \hline
        
        Aatresh et al.   & TCGA & 3 & LiverNet & Accuracy 97.72 & 5 fold & Architecture specific \\
        (2021)\cite{LiverNet} & 2240 train & ~ & ~ & Precision 97.72 & CV & for HCC liver cancer classification.\\
        ~ & 140 test & ~ & ~ & Recall 97.72 & ~ &  CBAM, ASPP block\\
        ~ & 224x224 & ~ & ~ & F1-score 97.72 & ~ & hypercolumn technique\\
        ~ & ~ & ~ & ~ & IOU$^{\mathrm{c}}$ 95.61 & ~ & less number of parameters\\ \hline

        Aatresh et al.  & TCGA & 3 & BreastNet & Accuracy 96.04 & 5 fold & Architecture specific \\
         (2021)\cite{LiverNet} & 2240 train & ~ & ~ & Precision 96.04 & CV & for Breast cancer classification.\\
        ~ & 140 test & ~ & ~ & Recall 96.04 & ~ &  CBAM block hypercolumn\\
        ~ & 224x224 & ~ & ~ & F1-score 96.04 & ~ & technique\\
        ~ & ~ & ~ & ~ & IOU 92.47 & ~ & less number of parameters\\ \hline

        Aatresh et al.   & KMC & 4 & LiverNet & Accuracy 90.93 & 5 fold & Architecture specific \\
        (2021)\cite{LiverNet} & 3210 train & ~ & ~ & Precision 90.93 & CV & for HCC Liver cancer classification.\\
        ~ & 272 test & ~ & ~ & Recall 90.93 & ~ &  CBAM, ASPP block\\
        ~ & 224x224 & ~ & ~ & F1-score 90.93 & ~ & hypercolumn technique\\
        ~ & ~ & ~ & ~ & IOU 83.6 & ~ & ~ \\ \hline

        Aatresh et al.  & KMC & 4 & BreastNet & Accuracy 88.41 & 5 fold & Architecture specific \\
        (2021) \cite{LiverNet} & 3210 train & ~ & ~ & Precision 88.41 & CV & for Breast cancer classification.\\
        ~ & 272 test & ~ & ~ & Recall 88.41 & ~ &  CBAM block hypercolumn\\
        ~ & 224x224 & ~ & ~ & F1-score 88.41 & ~ & technique\\
        ~ & ~ & ~ & ~ & IOU 79.73 & ~ & ~\\ \hline
        
        Bhattacharya et al.  & LC25000 & 2 & EfficientNeb4 & Accuracy 99.99 & 5 fold & AdBet-WOA, was applied \\
        (2023) \cite{BHATTACHARYA2023104692} & 5000 train & ~ & ~ &  F1-score 99.99 & CV & for feature selection, followed \\
        ~ & 500 test & ~ & ~ & ~ & ~ &  by classification using SVM\\
        ~ & 768x768 & ~ & ~ & & ~ & ~\\ \hline

        Talukder et al.  & LC25000 & 2 & Ensemble Learning & Accuracy 100.00 & 10 fold & Deep extracted features fed  \\
        (2022) \cite{TALUKDER2022117695} & 5000 train & ~ & ~ &  F1-score 100.00 & CV & to 6 different algorithms, \\
        ~ & 500 test & ~ & ~ & Precision 100.00 & ~ & ensemble learning\\
        ~ & 768x768 & ~ & ~ & Recall 100.00 & ~ & ~\\
        ~ & ~ & ~ & ~ & AUC 1.00 & ~ & ~\\ \hline

        Proposed  & TCGA & 3 & Pretrained & Accuracy 100.00 & 5 fold & Transfer learning, fine \\
        Hybrid & 3528 train & ~ & Hybrid & Sensitivity 100.00 & CV & tuning, deep classifier\\
        ResNet50 & 392 test & ~ & ResNet50 & Specificity 100.00 & ~ &  training of selective \\
        & 224x224 & ~ & ~ & F1-score 100.00 & ~ & top layer of base model \\
        ~ & ~ & ~ & ~ & AUC 1.00 & ~ & ~ \\ \hline

        Proposed  & KMC & 4 & Pretrained & Accuracy 96.71 & 5 fold & Transfer learning, fine \\
       Hybrid & 2725 train & ~ & Hybrid & Sensitivity 96.97 & CV & tuning, deep classifier\\
        EfficientNetb3 & 280 test & ~ & EfficientNetb3 & Specificity 98.85 & ~ &  training of selective \\
         & 224x224 & ~ & ~ & F1-score 96.71 & ~ & top layer of base model\\
        ~ & ~ & ~ & ~ & AUC 0.99 & ~ & ~\\ \hline

        Proposed  & LC25000 & 2 & Pretrained & Accuracy 100.00 & 5 fold & Transfer learning, fine \\
       Hybrid & 5000 train & ~ & Hybrid & Sensitivity 100.00 & CV & tuning, deep classifier\\
        EfficientNetb2 & 500 test & ~ & EfficientNetb2 & Specificity 100.00 & ~ &  training of selective \\
         & 768x768 & ~ & ~ & F1-score 100.00 & ~ & top layer of base model\\
        ~ & ~ & ~ & ~ & AUC 1.00 & ~ & ~\\ \hline

    \multicolumn{7}{l}{$^{\mathrm{a}}$MCC stands for Matthews's correlation coefficient. $^{\mathrm{b}}$CV stands for Cross Validation. $^{\mathrm{b}}$IOU stands for intersection over union}
    \end{tabular}
    \end{adjustbox}
\end{table*}

For 3-class patch-level classification on TCGA and 4-class patch-level classification on KMC datasets, the proposed hybrid methodology outperforms all the above-mentioned studies. Furthermore, the proposed model achieves better performance despite being trained on a comparatively smaller dataset and provides end-to-end deep learning-based solutions. The performance metrics also indicate the robustness of the proposed model over others. We have also demonstrated the performance of our methodology over colon histopathology images. Previously, Bhattacharya et al. \cite{BHATTACHARYA2023104692} and Talukder et al. \cite{TALUKDER2022117695} on the colon dataset claimed very large accuracies at the cost of large computational expenses. Whereas Hybrid models have achieved 100\% accuracy despite having a comparatively smaller number of parameters and end-to-end solution. Here as well, Hybrid models achieved 100\% accuracy.

There are some limitations to this study. The architectures based on ResNet50, EfficientNetb3, and DenseNet121 have approximately 25M, 12.5M, and 7M parameters, respectively. The high parameter count of these models makes them challenging to operate on low-end processors. To address this issue, we plan to utilize distillation methods or explore alternative approaches using low-rank deep learning models in future work, aiming to deliver highly accurate and lightweight solutions.

\section{Conclusions}
This study was focused on liver HCC classification using histopathology images. The primary dataset TCGA-LIHC was prepared through pre-processing steps involving patch extraction, color normalization and augmentation techniques. Another proprietary KMC dataset and a publically available COLON dataset were used for validation purposes.
Through this study, we proposed a deep learning-based hybrid architecture for the efficient and robust classification of liver HCC. The experimental results also show that the proposed hybrid model outperforms the base model and other state-of-the-art models by a significant margin. The best results on the TCGA dataset were obtained by a hybrid model with ResNet50 as the feature extractor, giving an accuracy of 100$\pm$0.00 whereas on the KMC dataset, the proposed hybrid model with EfficientNetb3 as a feature extractor achieves an accuracy of 96.71$\pm$0.68\%. Moreover, the proposed hybrid models provide outstanding results on the COLON dataset as well. The findings suggest that enhancing the classifier, optimizing its depth, and selectively fine-tuning the top layers of the feature extractor can significantly boost performance.

Despite having limitations of model parameters, since many labs are now equipped with high-end systems and because perfect predictions trump computational complexity in the medical domain due to the life-and-death situation, we believe a large number of parameters should not be a major issue. 

\section*{Acknowledgements}
This work was funded by the Science and Engineering Research Board,
India under Grant SERB- CRG/2021/005752. 

\ifCLASSOPTIONcaptionsoff
  \newpage
\fi
\bibliographystyle{ieeetr} 
\bibliography{Main}

\vspace{-10mm}
\end{document}